\documentclass{bmcart}

\usepackage[utf8]{inputenc} 
\usepackage[english]{babel}
\usepackage{graphicx}
\usepackage{subcaption}
\usepackage{framed}
\usepackage[normalem]{ulem}
\usepackage{amsmath}
\usepackage{amsthm}
\usepackage{amssymb}
\usepackage{amsfonts}
\usepackage{enumerate}
\usepackage{wrapfig}
\usepackage{float}
\usepackage{ragged2e}
\usepackage{url}
\usepackage{xcolor}
\usepackage{cleveref}
\usepackage[autostyle, english = american]{csquotes}
\MakeOuterQuote{"}
\usepackage{longtable}

\usepackage{cite}

\begin{document}

\begin{frontmatter}

\begin{fmbox}
\dochead{Research}


\title{Inference of Media Bias and Content Quality Using Natural-Language Processing}


\author[
   addressref={aff1},                   
   corref={aff1},                       
   email={zchao3@math.ucla.edu}   
]{\fnm{Zehan} \snm{Chao}}
\author[
   addressref={aff1},
   email={dmolitor@math.ucla.edu}
]{\fnm{Denali} \snm{Molitor}}
\author[
   addressref={aff1},
   email={deanna@math.ucla.edu}
]{\fnm{Deanna} \snm{Needell}},\!
\author[
   addressref={aff1,aff2},
   email={mason@math.ucla.edu}
]{\fnm{Mason A.} \snm{Porter}}
 

\address[id=aff1]{
  \orgname{Department of Mathematics, University of California, Los Angeles}, 
  \street{520 Portola Plaza},                     %
  \postcode{90095}                                
  \city{Los Angeles, CA},                              
  \cny{USA}                                    
}

\address[id=aff2]{
  \orgname{Santa Fe Institute}, 
  \street{1399 Hyde Park Road},                     %
  \postcode{87501}                                
  \city{Santa Fe, NM},                              
  \cny{USA}                                    
}


\begin{artnotes}
\end{artnotes}

\end{fmbox}


\begin{abstractbox}

\begin{abstract}
Media bias can significantly impact the formation and development of opinions and sentiments in a population. It is thus important to study the emergence and development of partisan media and political polarization. However, it is challenging to quantitatively infer the ideological positions of media outlets. In this paper, we present a quantitative framework to infer both political bias and content quality of media outlets from text, and we illustrate this framework with empirical experiments with real-world data. We apply a bidirectional long short-term memory (LSTM) neural network to a data set of more than 1 million tweets to generate a two-dimensional ideological-bias and content-quality measurement for each tweet. We then infer a ``media-bias chart'' of (bias, quality) coordinates for the media outlets by integrating the (bias, quality) measurements of the tweets of the media outlets. We also apply a variety of baseline machine-learning methods, such as a naive-Bayes method and a support-vector machine (SVM), to infer the bias and quality values for each tweet. All of these baseline approaches are based on a bag-of-words approach. We find that the LSTM-network approach has the best performance of the examined methods. Our results illustrate the importance of leveraging word order into machine-learning methods in text analysis. 

\end{abstract}




\end{abstractbox}
%

\end{frontmatter}




\section{Introduction}

Mass media is a fundamental part of modern society. Media outlets provide windows to the world and influence public knowledge, attitudes, and behavior \cite{gerber2009does}. They disseminate news and information, help educate the public, provide entertainment, and influence the spread of ideologies and opinions \cite{amedie2015impact}. However, media outlets have biases (including potentially very strong ones), and their influential societal roles make it important to examine such biases \cite{south2022,tien2020online}. Biased ideologies can impact people's choices (e.g., through their attitudes on topics like abortion \cite{rye2020pro}), their sharing behavior on social media \cite{cicchini2022news}, and more. Additionally, media can exacerbate political polarization by intensifying or even creating ideological ``echo chambers'' \cite{schober2016social, bail2018exposure} and enhancing so-called ``pernicious polarization'' \cite{mccoy2016polarized}, which divides societies into ``Us versus Them'' camps along a focal dimension of difference that overshadows other similarities and differences. 
There is a long history of quantitative studies of voting blocs and ideological biases of politicians \cite{pr1997,rice1927,sirovich2003}. Researchers have quantitatively examined the ideological biases of politicians in a variety of situations, including in social media \cite{huberman2008social, maynard2011automatic, digrazia2013more, preoctiuc2017beyond, lai2019stance, waller2020community} and in television interviews \cite{huls2011political}. 
Media outlets help broadcast the messages of public figures (such as politicians), which, in turn, influences public biases and opinions. Moreover, the quantitative study of the ideologies of politicians also helps guide investigations of the political biases of other entities, such as private citizens and media outlets, and one can estimate the ideological positions of individuals based on the media outlets with which they engage \cite{koz2021}. 

There are a variety of approaches to quantify ideological bias. It is common to use a liberal--conservative (i.e., ``Left--Right'') political spectrum as a one-dimensional (1D) spectrum when analyzing ideological biases~\cite{preoctiuc2017beyond, dixit2007political, panda2020covid, prior2013media}. This places these ideological biases in the context of common political polarities. A Left--Right dimension also arises in data-driven inference of ideological positions in multiple dimensions \cite{pr1997,boche2018new}. In the United States, it is traditional to place the Democratic political party on the Left and the Republican political party on the Right. Ideological views manifest in conversations and other ``digital footprints'' on social-media platforms~\cite{kitchener2022}, such as in posts by politicians on Twitter. For example, Anmol et al. \cite{panda2020covid} manually counted selected words in tweets that are related to COVID-19 and concluded that Republican politicians post more tweets that are related to business and the economy and that Democratic politicians concentrate more on public health. Xiao et al. \cite{patricia2022} examined political polarities in textual data from social-media platforms (specifically, from Twitter and Parler) and quantified such ideological biases on tweets from politicians and media outlets by assigning polarity scores to words, hashtags, and other objects (``tokens'') in social-media posts. Waller et al. \cite{waller2020community} introduced a multidimensional framework to summarize the ideological views, with a focus on traditional forms of identity (specifically, they considered age, gender, and political partisanship) of the posters and commenters, of Reddit posts around the time of the 2016 United States presidential election. Similarly, Gordon et al. \cite{gordon2020studying} argued that one cannot fully capture political bias using a single axis with two binary ideological extremes (such as Republican and Democrat); instead, one should use multidimensional approaches. Moreover, models that aim to analyze the content of media outlets should incorporate not only measures of outlet biases but also measures of outlet quality \cite{pennycook2019fighting, allen2020evaluating}. In the present paper, we use natural-language processing (NLP) \cite{ferr2020} of textual data from tweets by media outlets to infer (Left--Right, low--high) coordinates for these outlets, where the first dimension describes the political bias of a media outlet and the second dimension represents its quality. 

Neural-network models that are based on deep learning have been useful for many NLP tasks, including speech recognition \cite{mikolov2011strategies}, sentiment classification, answer selection, and textual entailment \cite{yin2017comparative}. By using deep neural networks instead of traditional machine-learning (ML) approaches (such as a naive-Bayes method), one can significantly improve the performance of tasks like text classification~\cite{sze2017efficient}. Moreover, neural networks that exploit input word sequences can produce more accurate results than methods that rely on a bag-of-words approach~\cite{iyyer2014political,salehinejad2017recent}. Recurrent neural networks (RNN) are a trendy deep-learning architecture to analyze sequential textual data. For example, Socher et al.~\cite{socher2013recursive} used an RNN to study binary sentiments (i.e., favorable or unfavorable) from a data set of movie reviews. In the present paper, we apply a specific type of RNN called a long short-term memory (LSTM) neural network to infer ideological and quality coordinates of media outlets based on the textual content of their tweets. We also compare the results of using an LSTM network to those from several traditional ML methods that have been used in previous sentiment analysis studies. These traditional approaches include a naive-Bayes method, a support-vector machine (SVM), an artificial neural network (ANN), a decision tree, and a random forest. All of these traditional approaches use a bag-of-words approach.

\begin{figure}[ht]
    \centering
	\begin{subfigure}[b]{0.95\linewidth}
	\includegraphics[width=\textwidth]
		{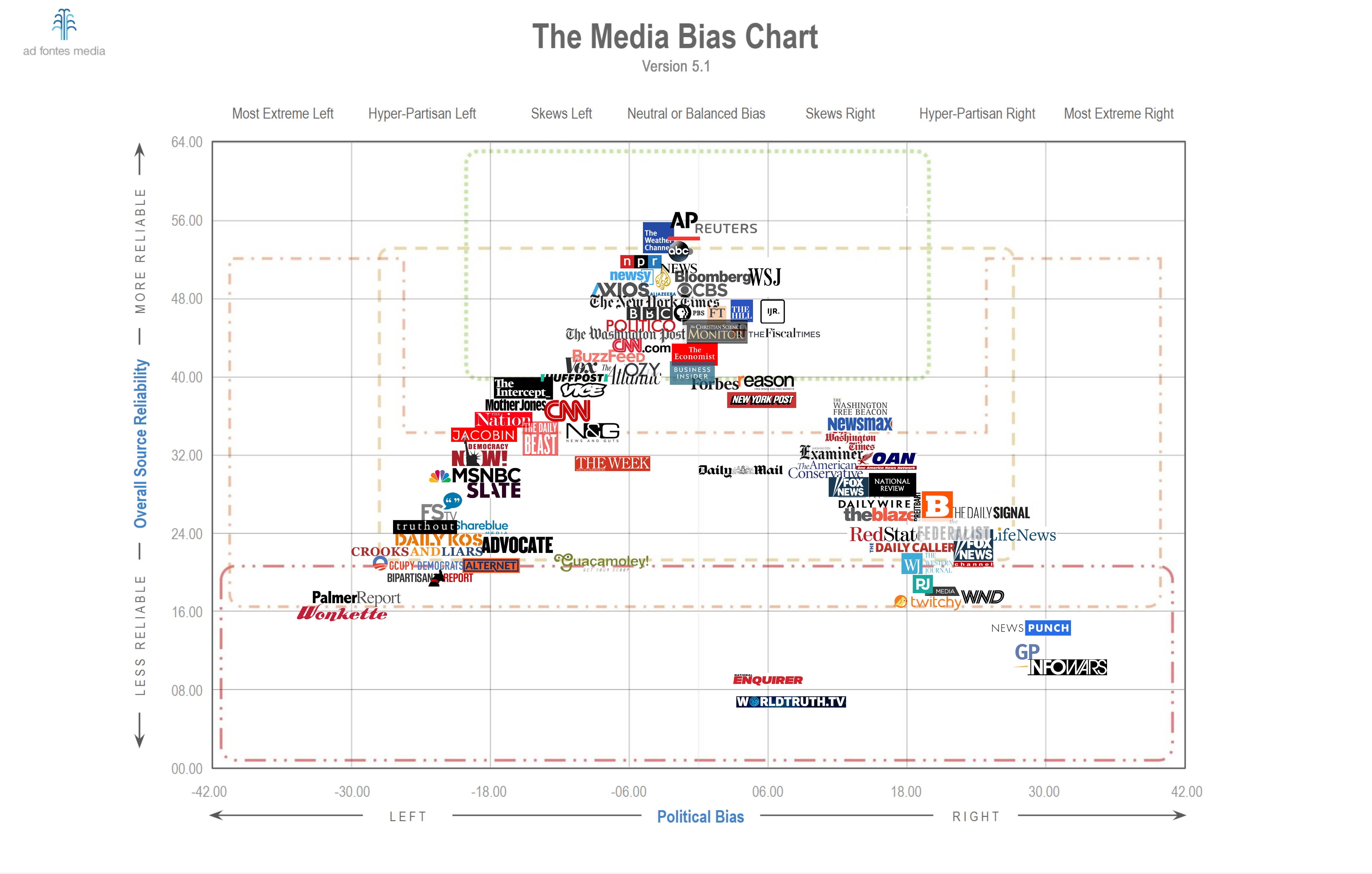}
	\end{subfigure}
	  \caption{The Ad Fontes Media-Bias Chart (version 5.1). When we use data from this bias chart (see \Cref{sec:Experiment}), we will normalize the bias scores to $[-1,1]$ and the quality scores to $[0,1]$.
{\footnotesize [We reproduce this figure, with permission that is granted by our purchase of a Standard License, from Ad Fontes. See {\tt https://adfontesmedia.com/copyright-and-usage-info/?utm\_source=StaticMBCPage}.}]
	  }
	  \label{fig-chart}
\end{figure}

To further motivate our work, consider the Ad Fontes Media-Bias Chart (AFMBC)~\cite{otero2019media}, which is a two-dimensional (2D) visualization (see Figure \ref{fig-chart}) of the political ideologies and content qualities of about 100 media outlets. The AFMBC shows the positions of media outlets with ideological biases along one axis and content qualities along the other axis. The AFMBC uses data from 1,818 online articles and 98 cable news shows, which were rated in 2019 by a politically balanced team of analysts \cite{otero2019media}. In the AFMBC, the bias and quality scores of each media outlet are the mean scores of each rated news item. Typically, 15--20 news items were used to evaluate a media outlet; at least three analysts rated each news item.

By necessity, the AFMBC uses a small number of items from each media outlet, although media outlets produce a wealth of content. 
By applying modern data-science techniques, one can leverage such abundant data through an algorithmic process of rating documents for both bias and quality.
For example, Widmer et al.~\cite{widmer2022media} applied penalized logistic regression and latent Dirichlet allocation (LDA) to a corpus of about 40,000 transcribed television episodes and examined the potential influence that national cable television can have on local newspapers. By measuring the textual similarities between the content of national television channels (specifically, Fox News, CNN, and MSNBC) and local newspaper content, they found that the content of local newspapers with more viewership of a given cable channel has greater textual similarity with the content of that cable channel than with it does with the other examined cable channels. They suggested the possibility that national cable television propagates slants and partisan biases to local newspapers and can thereby polarize local news content. In the present paper, we explore the potential to quantify the political ideologies and content qualities of media outlets based on the tweets that they posted in a specific time window. We use a data set from the Harvard GWU Libraries Dataverse \cite{DVN/2FIFLH_2017} of more than 30 million tweets from more than 4,000 news outlets. We then remove tweets that were not posted by media outlets in the AFMBC. This leaves about 1.4 million tweets, to which we apply ML techniques to infer bias and quality coordinates for each tweet. There is no absolute truth in the numerical values of a media outlet's bias and quality scores. Therefore, we evaluate the performance of the ML algorithms by comparing their outputs with the AFMBC. The 2D coordinates that we obtain from an algorithm can provide insight into ideological polarization and can serve as an input to opinion-dynamics models, such as the one in~\cite{brooks2020model} that incorporates media outlets.

\subsection{Our Contributions} 
\label{sec:overview}

\begin{figure}[ht]
    \centering
    \includegraphics[width=\textwidth]{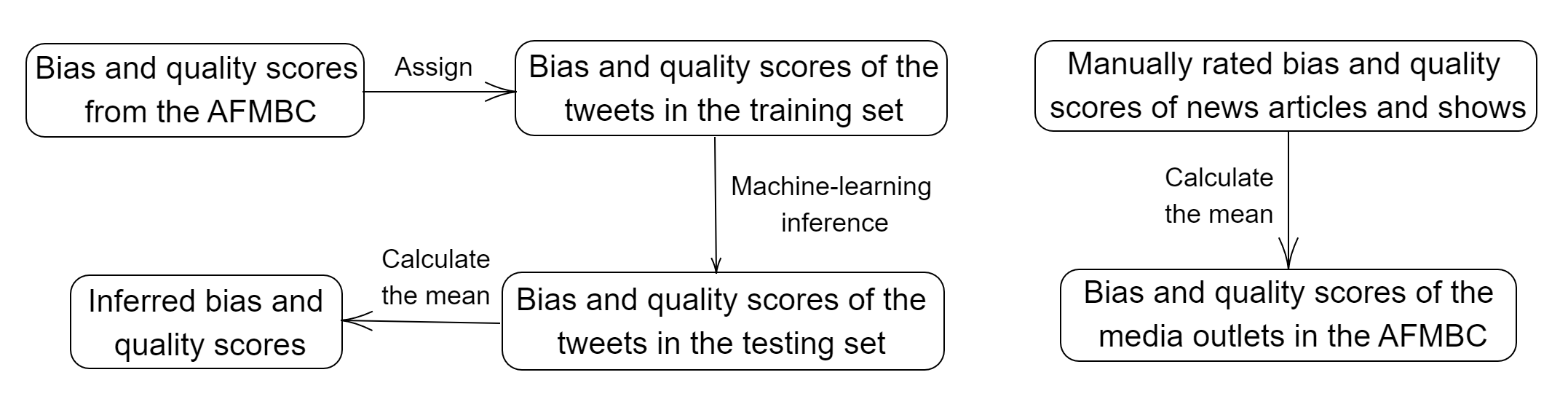}
    \caption{Flowcharts for inferring ideological biases and content qualities of the media outlets. We show (left) the workflow for our proposed approach and (right) the workflow for the AFMBC \cite{otero2019media}.
    }
    \label{flowchart}
\end{figure}

As of May 2020 (based on manual inspection), 65 of the media outlets in the AFMBC maintained active Twitter accounts. The other media outlets in the AFMBC either did not have a Twitter account or had a suspended account at that time. Between 4 August 2016 and 12 May 2020, these 65 accounts generated 1.4 million tweets; these tweets are tabulated in the George Washington University (GWU) Libraries Dataverse \cite{DVN/2FIFLH_2017} in the Harvard Dataverse. We use several existing supervised ML algorithms (a naive-Bayes method, an SVM, an ANN, a decision tree, a random forest, and an LSTM network) to infer the ideological biases and content qualities of each tweet. We then compute the bias and quality scores for each media outlet by calculating the means of the biases and qualities of their tweets. In \Cref{flowchart}, we show our workflow for inferring the bias and quality scores of the tweets and media outlets.

We observe a strong linear correlation between the manually rated political ideologies in the AFMBC (which plays the role of a ``ground truth'' in our study) and the political ideologies that are inferred by the LSTM network. We observe a similarly strong positive linear correlation between the AFMBC quality scores and the inferred quality scores. This suggests that algorithms can successfully produce reasonable (ideology, quality) scores for articles and other text from media outlets. We generate a 2D media-bias chart for the examined media outlets and compare it with the AFMBC. We then compare the results of an LSTM neural-network approach to the results of several traditional methods (a naive-Bayes method, an SVM, a decision tree, a random forest, and an ANN). We find that the LSTM approach (which considers word sequences) outperforms these baseline methods, which each use a bag-of-words approach.


\subsection{Organization of our Paper} 

Our paper proceeds as follows. In \Cref{sec: related model}, we discuss several ML methods that we use to infer the ideological biases and content qualities of tweets. In \Cref{sec:data}, we briefly describe the employed data, which includes the manually rated news articles and shows from Ad Fontes \cite{otero2019media} and the media-outlet tweets from the GWU Libraries Dataverse \cite{DVN/2FIFLH_2017}. In \Cref{sec:MBC}, we propose a scheme to construct a media-bias chart from the tweet content of media outlets. We also briefly describe our preprocessing of the tweets. In \Cref{sec:evaluate}, we introduce how we evaluate the performance of each ML method and compare the performance of different ML methods. In \Cref{sec:dis}, we conclude and discuss future work. In Appendix \ref{tab:media outlets}, we list the 65 examined media outlets and the number of tweets for each of them.



\section{Background and Related Work on NLP Methods}\label{sec: related model}

Because of the increased bounty and accessibility of machine-readable text \cite{vanloon_2022}, researchers have applied many supervised ML techniques to analyze and classify textual data \cite{kowsari2019text}. One can categorize supervised NLP methods into (1) sequential approaches (which account for word order) and (2) non-sequential approaches (which do not). Non-sequential methods transform a document into a bag of words before subsequent analysis \cite{cambria2014jumping}. Such methods were the typical type of approach in early NLP studies with ML techniques~\cite{bakliwal2011towards, kanakaraj2015performance, troussas2013sentiment}, and they are still the main approach for smaller data sets (e.g., ones with fewer than 5,000 data points~\cite{gopi2020classification}). Importantly, sequential models (e.g., RNNs \cite{mikolov2010recurrent}) transform a document into a sequential input and use associated contextual information when mapping from an input sequence to an output sequence \cite{graves2012long}. Therefore, we expect the inference of political ideologies to be more effective with a sequential approach than with a bag-of-words approach~ \cite{socher2013recursive}.

In this section, we briefly discuss several non-sequential approaches (which we employ as baseline methods) and a sequential approach that we use to infer a media-bias chart from the tweets of media outlets. In \Cref{notation}, we summarize our key notation.

\begin{table}[H]
    \centering
    \begin{tabular}{|c|l|l|}\hline
    symbol & explanation & example or further information \cr \hline \hline
        $\mathbf{X}$ &  an entire data set or a training set & example: the entire set of tweets \cr 
        \hline
        $\mathbf{x}$ \,or\, $\mathbf{x}_i$ &  an input feature vector & example: one tweet \cr 
        \hline
        $x_j$ & the $j$th entry of a
        vector $\mathbf{x}$ & example: one word \cr 
        \hline
        $y_k$ & the label of the $k$th input & example: the label of the $k$th tweet \cr 
        \hline
        $\mathbf{w}$ & a vector of weights & we use these in the SVM and the neural networks \cr 
        \hline
        $b$ & the bias term in $\mathbf{w} \cdot \mathbf{x} + b$ & we use these in the SVM and the neural networks \cr 
        \hline
        $c$ & the index of a label & it ranges from $1$ to $n$ when there are $n$ classes in total \cr 
        \hline
    \end{tabular}
    \caption{Key notation in our paper.
    }
    \label{notation}
\end{table}


\subsection{Naive-Bayes Method}

A naive-Bayes method is a simple approach that has been used successfully for text categorization \cite{troussas2013sentiment,go2009twitter,manning1999foundations}, which is the common NLP task of assigning each text document in a corpus to a category $c \in \{1,\ldots ,n\}$.

We start with Bayes' rule
\begin{equation*}
	\mathbb{P}(y_i=c \mid \mathbf{x}_i)=\frac{\mathbb{P}(y_i=c) \mathbb{P}(\mathbf{x}_i \mid y_i=c)}{\mathbb{P}(\mathbf{x}_i)}\, .
\end{equation*}
Using the chain rule, we compute the probability that the current item $i$ (for example, the $i$th tweet), with feature vector $\mathbf{x}_i = [x_{i_1}, \ldots, x_{i_n}]^T$ (for example, the sequence of words in the $i$th tweet), is in category $c$. This yields
\begin{align*}
	&\mathbb{P}(y_i=c \mid x_{i_1}\text{,} \ldots, x_{i_n})= \mathbb{P}(x_{i_1} \mid x_{i_2}, \ldots, x_{i_n}, y_i=c)\,\times \cdots \\ 
	&\qquad \qquad \qquad \qquad \qquad \quad \times \,\mathbb{P}(x_{i_{n-1}}\mid x_{i_n}, y_i=c)\, \times \,\mathbb{P}(x_{i_n} \mid y_i=c)\, \times \,\mathbb{P}(y_i=c)\,.
\end{align*}

The word ``naive'' appears in the method's name because a naive-Bayes approach assumes that all words have independent probabilities of appearing in a document. That is,
\begin{align*}
	\mathbb{P}(x_{i_1} \mid x_{i_2}\text{,} \ldots, x_{i_n}, y_i=c) = \mathbb{P}(x_{i_1} \mid y_i=c)\,.
\end{align*}	

In a naive-Bayes approach, one needs both $\mathbb{P}(y_i=c)$ and $\mathbb{P}(x_j \mid y_i=c)$ to compute the output probability $\mathbb{P}(y_i=c \mid x_{i_1}\text{,} \ldots, x_{i_n})$. One substitutes the probability $\mathbb{P}(y_i=c)$ for the relative frequency of class $c$ in the training set. We obtain the conditional probability $\mathbb{P}(x_j \mid y_i=c)$ using maximum a posteriori (MAP) estimation \cite{zhang2004optimality}.


\subsection{Support-Vector Machines (SVMs)}

SVMs have been employed often in classification tasks \cite{bhavsar2012review, sheykhmousa2020support, toledo2019support}. For example, Go et al. \cite{go2009twitter} used SVMs for sentiment classification of Twitter data. Gopi et al.~\cite{gopi2020classification} used an SVM approach for tweet classification using positive and negative opinions. 

In a traditional SVM, one is given a training data set $\{\mathbf{x}_1, \ldots, \mathbf{x}_n\}$ and seeks binary class labels, which we denote by $+1$ and $-1$. One seeks the maximum-margin hyperplane that separates the data points of those two different classes. One attempts to minimize the hinge loss
\begin{equation} \label{hinge}
	\left[\frac{1}{n} \sum_{i=1}^{n} \max \{0,1-y_{i}\left(\mathbf{w}\cdot \mathbf{x}_{i}+b\right)\}\right]+\lambda\|\mathbf{w}\|^{2}\, ,
\end{equation}
where the parameter $\lambda >0$ determines the trade-off between the margin size and the labeling accuracy, which is equal to the number of correctly assigned labels divided by the number of elements in the training set. The loss function \eqref{hinge} is convex, so one can use a common convex-optimization approach (e.g., gradient descent) to successfully minimize it \cite{rennie2005loss}.

The original SVM setting is directly applicable only to binary classification. For classification problems with three or more labels, one needs to split the classification task into multiple binary-classification tasks. 


\subsection{Decision Trees and Random Forests}

A decision tree is a flowchart-like structure in which each node $t$ of the tree splits the flow of a procedure into two sets of classes based on the information gain $\operatorname{I}(t)=H(\mathbf{X})-H(\mathbf{X}\mid t)$, where 
\begin{equation*}
	H(\mathbf{X}) = - \sum_c \mathbb{P}(y_i=c)\log_2(\mathbb{P}(y_i=c)) 
\end{equation*}
is the entropy of a data set and
\begin{equation*}
	H(\mathbf{X}\mid t) = \sum_{d \in s(t)} p(d) H(\mathbf{X} \mid d)
\end{equation*}
is the conditional entropy of a data set after the split at node $t$, where $s(t)$ denotes the set of all possible splits at $t$ with respect to one variable.

\begin{figure}[ht]
    \centering
    \includegraphics[width = 0.4\textwidth]{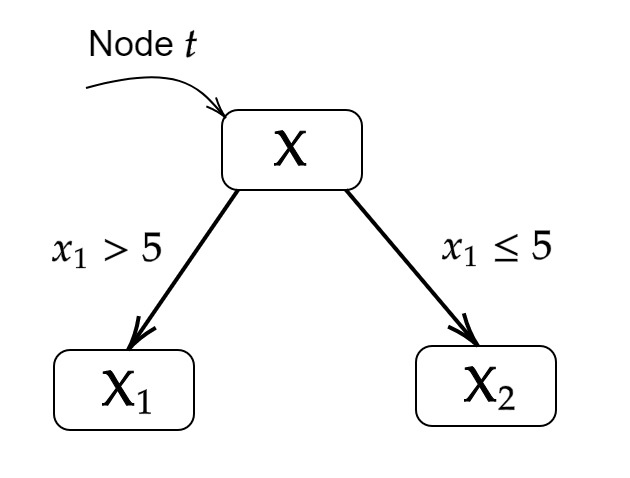}
    \caption{A schematic illustration of a simple decision tree. This example uses $x_1>5$ as the criterion to split the source data $\mathbf{X}$ into $\mathbf{X}_1$ and $\mathbf{X}_2$. 
    }
    \label{fig:dtdemo}
\end{figure}

In \Cref{fig:dtdemo}, we give a schematic illustration of splitting a source data set. In this example, the full data set is $\mathbf{X}$, the subsets $\mathbf{X}_1$ and $\mathbf{X}_2$ are disjoint, and $\mathbf{X}=\mathbf{X}_1 \cup \,\mathbf{X}_2$. The entropy of the data set after the split at node $t$ is
\begin{equation*}
	H(\mathbf{X}\mid t) = \frac{|\mathbf{X}_1|}{|\mathbf{X}|} H(\mathbf{X}_1)+\frac{|\mathbf{X}_2|}{|\mathbf{X}|} H(\mathbf{X}_2)\,.
\end{equation*}	
The root node is always split into two new nodes (which are called ``children''), and we further split any new node that has data with multiple labels. We continue the splitting process recursively (i.e., following a tree structure), using the criterion of maximum information gain at each node until we obtain a set of leaf nodes in which each node has data with a single label. See Breiman et al.~\cite{breiman2017classification} for other splitting strategies and stopping conditions for decision trees.

A random-forest classification approach uses a set of decision trees. It selects a random sample from a training set using some probability distribution and then fits trees to these samples \cite{breiman2017classification}. A random-forest classifier consists of $N$ trees, where one specifies the value $N$. We use $N = 100$ to balance efficiency and performance and to avoid overfitting. To classify a new data set, we pass each sample of that data set to each of the $N$ trees. The forest chooses a class with the most votes of the $N$ possible votes; if there is a tie for the largest vote total, it uniformly randomly selects one of the classes with the most votes. In one example of sentiment analysis using a random-forest classifier, Bilal et al.~\cite{bilal2016sentiment} identified positive, negative, and neutral sentiments in a set of documents.


\subsection{Artificial Neural Networks (ANNs)}

Artificial neural networks (ANNs) are popular classifiers that have been used for various text-classification problems \cite{zharmagambetov2015sentiment}. We use a fully-connected (i.e., ``dense'') feedforward ANN to infer media bias and back-propagation (BP) to train this ANN. Because of the feedforward structure, the nodes are not part of any cycles. BP is an iterative gradient-based algorithm that we use to minimize the mean-squared error between the actual output and the desired output. See Schmidhuber~\cite{schmidhuber2015deep} for more details about ANNs and training methods for them. ANNs have been used for a variety of tasks in sentiment analysis  \cite{ghiassi2013twitter,sharma2012artificial,sharma2012document}. Zharmagambetov and Pak~\cite{zharmagambetov2015sentiment} used an ANN and a word-embedding model to classify a data set of movie reviews with positive and negative sentiments.


\subsection{Long Short-Term Memory (LSTM) Neural Networks} \label{sec:relatedLSTM}

\begin{figure}[h!]
    \centering
    \includegraphics[width = \textwidth]{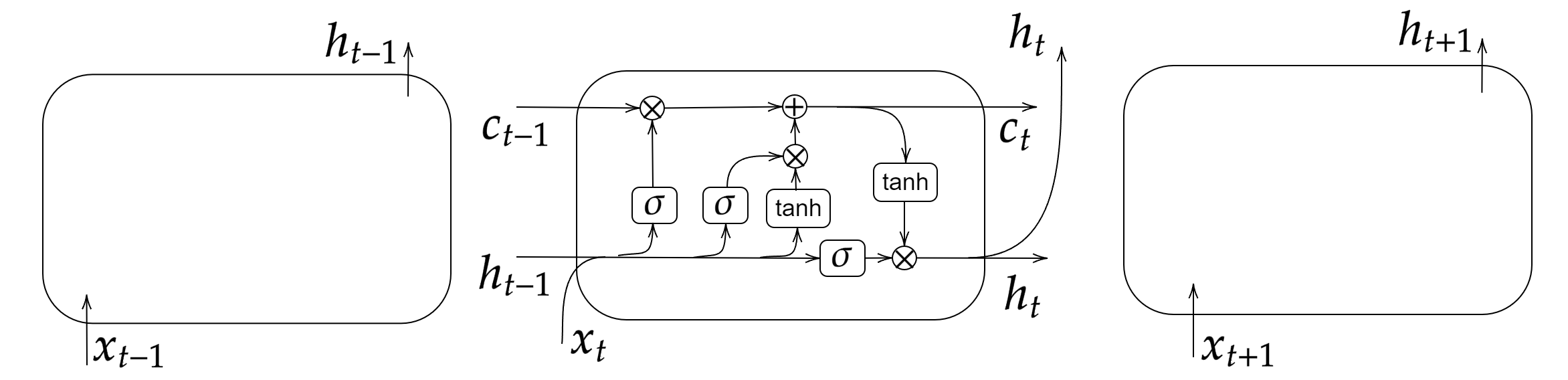}
    \caption{A schematic illustration of a memory cell in an LSTM neural network. The center panel is the $t$th cell, on which we focus.
        The quantity $x_t$ is the input (e.g., a sequence of integers that represents all words in one document), $c_t$ is the ``memory-cell state'' that is passed to the next cell, and the ``hidden state'' $h_t$ is also passed to the next cell. The function $\sigma$ is a sigmoid activation function, and the hyperbolic tangent \text{tanh} is another activation function.  
    In \Cref{sec:lstm}, we precisely specify the functions of the gates in each cell.
    }
    \label{LSTMexample}
\end{figure}

An LSTM neural network is one type of RNN architecture. Unlike a traditional RNN, an LSTM network has a ``forget'' gate that determines whether or not to pass information from one memory cell to the next memory cell. LSTM networks have achieved good performance on various NLP tasks, including machine translation, next-word inference, and binary sentiment classification~\cite{wang2016attention, young2018recent, tang2015target}. In \Cref{LSTMexample}, we show the structure of a single LSTM cell. The scalar $c$ denotes the ``memory'' that is passed between each cell. In the center panel (which indicates the $t$th cell), the sigmoid activation function $\sigma$ on the left is a forget gate. When $\sigma$ outputs $0$, the previous memory $c_{t-1}$ is ``forgotten'' by multiplying it by $0$; when $\sigma$ outputs $1$, one uses $c_{t-1}$ and passes it to the next cell. The middle sigmoid activation function $\sigma$ is called the input gate, which decides the input value for updating the memory. The right sigmoid activation function $\sigma$ is called the ``output gate'' and determines the output value. The hyperbolic tangent is another activation function, and the output is $h_t$. In \Cref{sec:lstm}, we will explain the architecture of the bidirectional LSTM network model that we use to infer the ideological biases and content qualities of the media outlets.

\section{Data Sets} \label{sec:data}

We use tweets from media outlets to examine the ideological biases and content qualities of those outlets. Tweets by media outlets are more readily available and larger in number than shows and articles from these outlets. We use data from the \textsc{NewsOutletTweet} data set of media tweets from Littman et al.~\cite{DVN/2FIFLH_2017}. This data set has a list of tweet IDs from the Twitter accounts of 9,636 news outlets. The tweets from these media outlets were collected between 4 August 2016 and 12 May 2020. The tweet IDs are 18-digit integers that provide access (with \texttt{twitter.com/} as the prefix) to the corresponding tweets. We use the Hydrator application \cite{hydrator2020} to extract the content of all 39,695,156 tweets in \textsc{NewsOutletTweet}; we refer to these tweets as the ``hydrated \textsc{NewsOutletTweet}'' data set. For each of the 102 media outlets in the AFMBC, we manually search for their official Twitter account and check if it is in the account list in \textsc{NewsOutletTweet}. If one media outlet has multiple official Twitter accounts --- e.g., \emph{The New York Times} has several Twitter accounts, such as \texttt{@NYTSports} and \texttt{@nytopinion} --- we manually determine its primary Twitter account and keep only this account. (For example, we use \texttt{@nytimes} for \emph{The New York Times}.) This yields 65 media outlets that are part of both the AFMBC and \textsc{NewsOutletTweet}. We list these media outlets and the associated Twitter accounts in \Cref{tab:media outlets}.

We select the tweets in the hydrated \textsc{NewsOutletTweet} data set that were posted by the 65 media outlets. There are 1,417,030 such tweets in total. Except for Bloomberg (for which there are 127 tweets), the numbers of tweets of the media outlets range from about 1,000 to about 100,000. We obtain bias and quality scores for each media outlet using the bias and quality scores of articles and shows in the \textsc{MediaSourceRatings} data set~\cite{otero2019media}. For each article, the ideological bias lies in one of seven categories: most extreme Left, hyperpartisan Left, skews Left, neutral, skews Right, hyperpartisan Right, and most extreme Right. Each category spans 12 rating units (except for the neutral category, which spans 13 units from $-6$ to $6$), so the numerical scale consists of all integers between $-42$ and $+42$. The seven categories of bias were defined by analysts, and the 12 units within each category allow nuanced distinctions in the amount of bias~\cite{otero2019media}. The overall reliability of each article by each media outlet was divided by the analysts into eight categories, with eight units each. This yields a numerical scale that consists of all integers between $1$ (the least reliable) and $64$ (the most reliable)~\cite{otero2019media}. The number of rating units was selected because it is convenient for the aspect ratio of visual displays.

Each media outlet in the AFMBC has between 15 and 25 reviewed articles and shows (in total, including both types of media) in the \textsc{MediaSourceRatings} data set. For each media outlet, we use the mean of the bias scores and the mean of the quality scores as representative quality and bias scores.


\section{Generation of a Media-Bias Chart}
\label{sec:MBC}

\begin{figure}[h!]
    \centering
    \includegraphics[width = 0.8\textwidth]{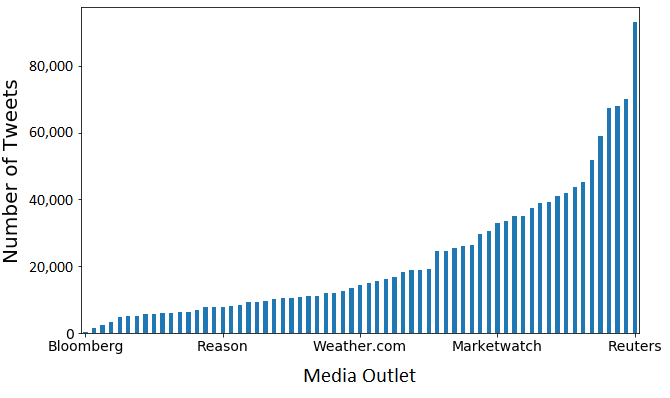}
    \caption{The number of tweets of the Twitter accounts of the 65 examined media outlets. Because of space considerations, we only show a few outlet names in this plot. For the complete list of media outlets, see \Cref{list_of_media}.
    }
    \label{TweetPerMedia}
\end{figure}

We preprocess the tweet data from \cite{DVN/2FIFLH_2017} and input it into an LSTM neural network to generate a bias score and a quality score for each tweet.\footnote{Our code for filtering the tweets from \textsc{NewsOutletTweet} and applying ML algorithms to this data set is available at \url{https://gitlab.com/zchao3/MediaSentiment.git}.} Using the (bias, quality) scores of the tweets (in the testing set), we generate a media-bias chart --- with coordinates for both ideological bias and content quality --- for the media outlets and then interpret it. The numbers of tweets of the media outlets range between 127 (for Bloomberg) and 93,259 (for Reuters); see \Cref{TweetPerMedia}. For our data preprocessing, we select 10,000 tweets uniformly at random with replacement (so we do bootstrap sampling, and the sampled tweets can contain duplicates, especially for media outlets with smaller numbers of tweets) for each media outlet. With this procedure, our sample data set has 65 $\times$ 10,000 tweets. The training set (a uniformly random subset) consists of a fixed proportion of the sampled tweets, to which we assign bias and quality scores that are equal to the bias and quality scores of the corresponding media outlets in the AFMBC. The media outlets thereby contribute equally to our training set regardless of how often they tweeted.


\subsection{Text Preprocessing}\label{sec:text process}

We apply the following text-preprocessing scheme to our data set. We remove all stop words (``and'', ``you'', ``to'', and so on) in the built-in list of stop words in the Python package \textsc{nltk}, and we then remove all hyperlinks. We build a vocabulary using the 5,000 words (where we treat hashtags as words) with the highest frequencies in the sampled data set using the \textsc{CountVectorizer} function in the Python package \textsc{sklearn}. Each word that remains in a tweet is a token. We convert all words that are not in the vocabulary into \text{``OOV''} tokens and add \text{``PAD''} tokens at the end of each tweet so that all tweets have the same length (i.e., the same number of tokens) as the longest tweet. 


\subsection{Bidirectional LSTM Neural Network} \label{sec:lstm} 

We input the processed tweets (which are sequences of tokens) into a bidirectional LSTM neural network to infer their bias and quality scores. The LSTM network that we use has three layers, which have different purposes. Each layer transforms an input sequence into another sequence of scalars or into a sequence of vectors. The first layer is a fully-connected word-embedding layer \cite{gulli2017deep} that transforms each word into a vector of a user-selected length. The second layer is a bidirectional LSTM layer that consists of a forward LSTM and a backward LSTM. The number of memory cells (see \Cref{sec:relatedLSTM}) in each LSTM is equal to the length of the input sequence. Each LSTM layer transforms an input sequence of vectors into a sequence of scalars. For each cell in the LSTM layer, we apply the following transition functions:
\begin{equation*}
	\begin{aligned}
	f_t &= \sigma(\textbf{w}_f\cdot[h_{t-1},\textbf{x}_t]+b_f) \,,\\
	i_t &= \sigma(\textbf{w}_i\cdot[h_{t-1},\textbf{x}_t]+b_i) \,,\\
	\tilde{c}_t &= \tanh(\textbf{w}_c\cdot[h_{t-1},\textbf{x}_t]+b_c) \,, \\
	c_t &= f_t c_{t-1} + i_t \tilde{c}_t \,,\\
	o_t &= \sigma(\textbf{w}_o\cdot[h_{t-1},\textbf{x}_t]+b_o) \,,\\
	h_t &= o_t  \tanh(c_t) \,.
	\end{aligned}
\end{equation*}
Following common practice, we refer to $f_t$, $i_t$, $o_t$, $c_t$, and $h_t$ as a forget gate, an input gate, an output gate, a memory-cell state, and a hidden state, respectively \cite{rao2018lstm, yu2019review, zhao2017lstm}. (The difference between a ``state'' and a ``gate'' is that the value of the former is passed to the next cell, but that is not the case for the latter.) The sigmoid function $\sigma(x)=\frac{1}{1+e^{-x}}$ and $\tanh$ function are the activation functions. Finally, we append a fully-connected concatenation layer \cite{Huang_2017_CVPR}
that transforms the outputs from the bidirectional LSTM into a pair of scalars in $[-1,1] \times [0,1]$ that encode a tweet's bias score and quality score. Both the embedding layer and the concatenation layer use a matrix--vector product and a rectified linear unit (ReLU) as an activation function. 
We use the standard ReLU activation function \cite{agarap2018deep} 
\begin{align*}
	f(x) 	&= \begin{cases}
				0  \,,& x \leq 0 \\
				x\,, &  x>0
			\end{cases} \notag \\
		&= \max \{0, x\} = x \, 1_{x>0} \,.
\end{align*}

\begin{figure}[h!]
    \centering
    \includegraphics[width = 0.5\textwidth]{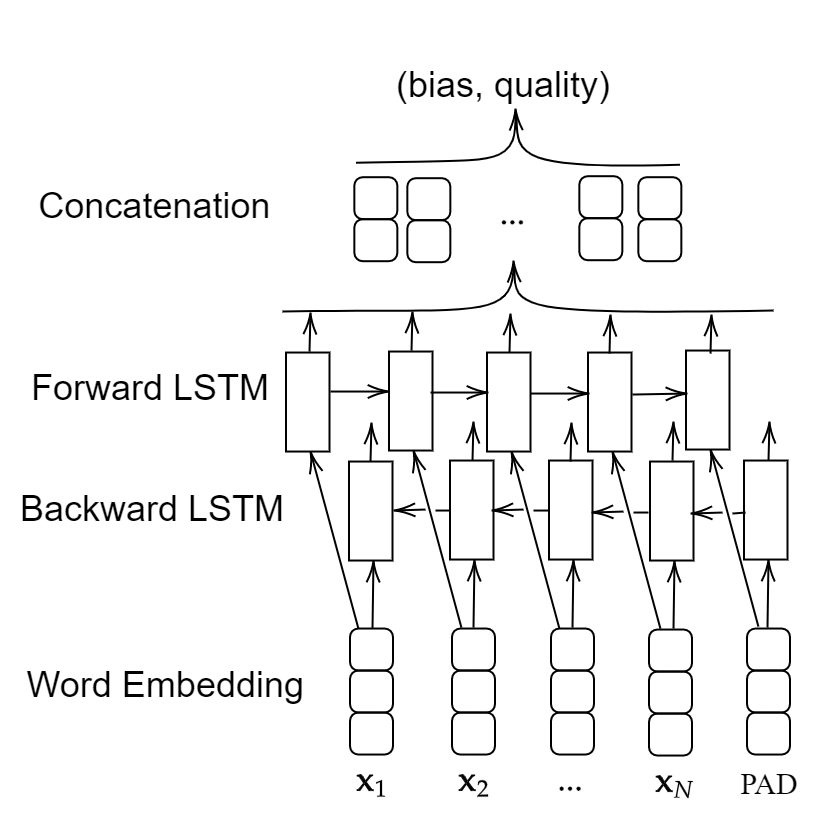}
\caption{The structure of our bidirectional LSTM. The concatenation layer is a fully-connected layer that takes output sequences from both a forward LSTM and a backward LSTM as input. The output of the concatenation layer is a pair of scalars for the inferred bias and quality scores. The term ``PAD'' indicates the padding tokens that we append to a tweet so that all tweets have the same length (i.e., the same number of tokens).
    }
    \label{BLSTM}
\end{figure}

In \Cref{BLSTM}, we show a schematic illustration of the bidirectional LSTM architecture and the workflow in our computational experiments. We initialize all parameters --- including the parameters in the embedding layer, the concatenation layer, and the weight vectors $\textbf{w}_f$, $\textbf{w}_i$, $\textbf{w}_C$, and $\textbf{w}_o$ --- with independent uniform random real numbers in $[0,1]$. As a loss function, we use the mean-square error (MSE) $\mathrm{MSE}=\frac{1}{n} \sum_{i=1}^{n}\left(y_{i}-\hat{y}_{i}\right)^{2}$, where $y_i$ denotes the $i$th training label and $\hat{y}_i$ denotes the $i$th inferred label, and we train the weight vectors with the Adams optimizer~\cite{kingma2014adam} to minimize the loss function.

\subsection{Training and Results} \label{sec:train}

\begin{figure}[h!]
    \centering
    \includegraphics
    [trim=120 40 90 80,clip,width = \textwidth]
   {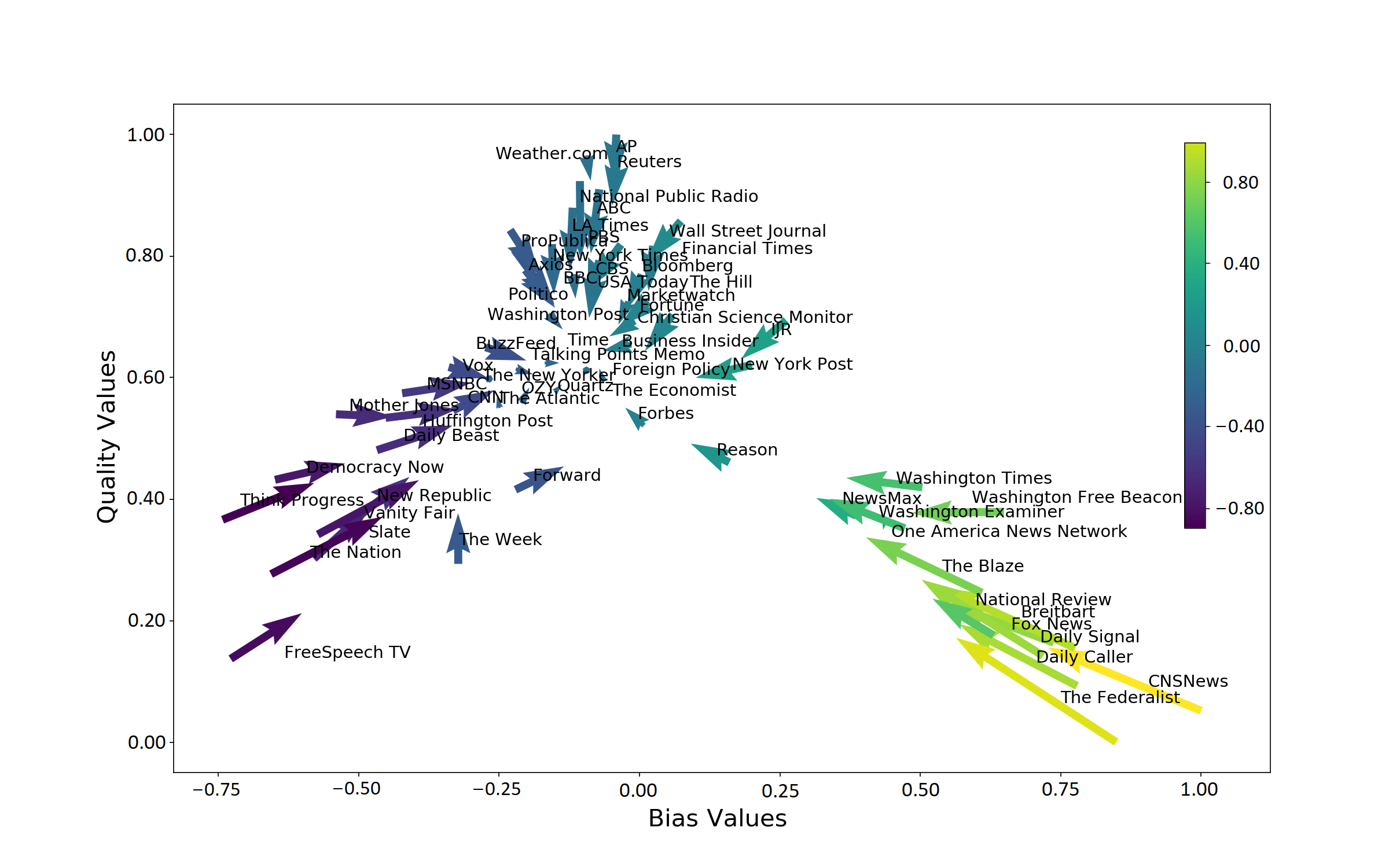}
    \caption{A comparison of the (bias, quality) scores from the AFMBC to the (bias, quality) scores that we obtain using an LSTM network. Each arrow starts from a media outlet's coordinates in the AFMBC and terminates at the output coordinates from the LSTM network. The color of each arrow indicates the media outlet's position on the Left--Right political spectrum (i.e., its ideological-bias score) in the AFMBC.
    }
    \label{fig:arrows}
\end{figure}

We use a tweet-level split of testing and training data. We split the 650,000 tweets uniformly at random with 80\% of them in the training set and 20\% in the testing set. We apply the trained model to the testing tweets and assign bias scores and quality scores to each of these tweets. We group the testing set by media outlet, and we then compute the mean bias score and mean quality score of each group and assign this pair of scores to the corresponding media outlet. We normalize the bias scores to $[-1,1]$, and we normalize the quality scores to $[0,1]$. In \Cref{fig:arrows}, we plot these scores along with the normalized bias and quality scores from the AFMBC.

\section{Evaluation of the LSTM Network's Performance}\label{sec:evaluate}

We examine the performance of inferring a 2D media-bias chart of (bias, quality) coordinates for each media outlet using a variety of approaches (see \Cref{sec: related model}).  As we will see, the LSTM network (which incorporates sequential data) outperforms the other approaches, which use only non-sequential information.

In the output of each method, we observe that many tweets do not have a strong ideological bias. This is the case even for many tweets from media outlets with a strong ideological bias (e.g., with an absolute value of at least $20$ in the interval $[-42,42]$) in the AFMBC. Therefore, it is sensible that the mean bias scores of the tweets of the media outlets tend to be closer to the center of the ideological spectrum than their bias scores from the AFMBC (see \Cref{fig:arrows}). 

We calculate Pearson correlation coefficients, which are invariant under scaling and shifting of its arguments, to examine linear correlations between the inferred bias and quality scores and the AFMBC bias and quality scores. We use these coefficients to evaluate the performance of each method; a large Pearson correlation coefficient suggests that a method yields a spectrum with a reasonable ordering of the scores. For example, a large Pearson correlation coefficient for the bias score indicates a good performance at inferring the relative locations of media outlets on a Left--Right ideological spectrum. The Pearson correlation coefficient of two vectors $\mathbf{y}$ and $\hat{\mathbf{y}}$ is
\begin{equation*}
	\rho(\mathbf{y}, \hat{\mathbf{y}}) =\frac{\mathbb{E}\left[\left(\mathbf{y}-\mu_{\mathbf{y}}\right)\left(\hat{\mathbf{y}}-\mu_{\hat{\mathbf{y}}}\right)\right]}{\sigma_{\mathbf{y}} \sigma_{\hat{\mathbf{y}}}}\,,
\end{equation*}
where $\mathbf{y}$ consists of the media-outlet bias or quality coordinates from the AFMBC and $\hat{\mathbf{y}}$ consists of the algorithmically inferred media-outlet coordinates from tweets.


\subsection{Computational Experiments}\label{sec:Experiment}

In addition to the LSTM-network approach on which we focus, we examine results from five other methods: a naive-Bayes (NB) method, a support-vector machine (SVM), a decision tree (DT), a random forest (RF), and an artificial neural network (ANN) with fully-connected layers. 

We compare the (bias, quality) coordinates that we obtain for each media outlet from these methods. For each tweet, we remove the stop words and hyperlinks (see \Cref{sec:text process}), and we use the 5,000 most-frequent words (see \Cref{sec:lstm}) to build a vocabulary. Our results vary slightly (with the Pearson correlation differing within about $\pm 0.03$ in our experiments) when we use vocabularies with 2,000 and 10,000 words. We apply an 80--20 train--test split of the data (see Section \ref{sec:MBC}) and generate (bias, quality) coordinates for each media outlet by calculating the mean of each method's output for the tweets from the same media outlet. We also test each method with train--test splits that we base on the media outlets themselves. In this media-outlet split, we select 52 media outlets (i.e., 80\% of them) uniformly at random. We train each method using all of the tweets from these 52 media outlets and apply the trained method to the tweets of the remaining 13 media outlets to infer (bias, quality) coordinates of all remaining media outlets.


\subsection{Results}

We now compare and discuss our results from the various approaches for inferring (bias, quality) coordinates. We consider both a media-level split (see \Cref{sec:Experiment}) and a tweet-level split (see \Cref{sec:train}).

For the media-outlet split, we uniformly randomly split the media outlets into five groups of 13 media outlets each, and we apply 5-fold cross-validation to generate the coordinates for all media outlets. That is, using each of the five groups as a withheld testing group, we train each method with the tweets from the other four groups and then evaluate the method with the tweets from the withheld group. We then compute the media bias and quality scores by calculating means of the scores of the tweets (see \Cref{sec:train}) and compute the Pearson correlation coefficients from these results. We show the mean Pearson correlation (which we average over five trials) for the media-outlet split in \Cref{tab:corr1}.

\begin{table}[H]
    \centering
    \begin{tabular}{|c|c|c|}\hline 
Method               &  Bias-Score Correlation & Quality-Score Correlation \\ \hline \hline
NB &  0.662 & 0.713\\ \hline
SVM & 0.652 & 0.736\\ \hline
DT & 0.665 & 0.684\\ \hline
RF & 0.780 & 0.779\\ \hline
MLP & 0.809 & 0.811\\ \hline
LSTM & \textbf{0.832} & \textbf{0.825}\\ \hline
    \end{tabular}
    \vspace{0.15cm}
    \caption{The Pearson correlations between the bias and quality scores in the AFMBC and the corresponding scores that we obtain from ML methods using a media-outlet split. For each method, the standard deviation of the correlations from the five different train--test splits in our 5-fold cross validation is smaller than 0.01. We show the means of the five different train--test splits for both the bias-score and quality-score correlations. We show the best results in bold.
    }
    \label{tab:corr1}
\end{table}

We also use 5-fold cross validation for the tweet-level split. We uniformly randomly split all tweets into five equal-sized groups. Using each of the five groups as a withheld testing group, we train each method using the other four groups and then evaluate the method using the withheld group. We show the mean Pearson correlation (which we average over five different train--test splits) for the tweet-level split in \Cref{tab:corr2}.

\begin{table}[H]
    \centering
    \begin{tabular}{|c|c|c|}\hline 
Method               &  Bias-Score Correlation & Quality-Score Correlation \\ \hline \hline
NB &  0.861 & 0.805\\ \hline
SVM & 0.909 & 0.898\\ \hline
DT & 0.856 & 0.861\\ \hline
RF & 0.925 & 0.907\\ \hline
MLP & 0.916 & 0.949\\ \hline
LSTM & \textbf{0.977} & \textbf{0.964}\\ \hline
    \end{tabular}
    \vspace{0.15cm}
    \caption{The Pearson correlations between the bias and quality scores in the AFMBC and the corresponding scores that we obtain from ML methods using a tweet-level split. 
    For each method, the standard deviation of the correlations from the five different train--test splits in our 5-fold cross validation is smaller than 0.01. We show the means of the five different train--test splits for both the bias-score and quality-score correlations. We show the best results in bold.
    }
    \label{tab:corr2}
\end{table}

From \Cref{tab:corr1} and \Cref{tab:corr2}, we see that the LSTM-network approach that incorporates word sequences in its input performs better than the other five methods, which all use a bag-of-words approach. This demonstrates that it is advantageous to account for sequential information. We also observe that all methods perform better on the tweet-level split than they do on the media-outlet split. One possible reason is that different media outlets use different word choices for different tweets. Moreover, media outlets with similar ideological biases and quality scores on the AFMBC may tend to post about different topics. For example, \texttt{Weather.com} and \texttt{The Economist} have the same ideological-bias score (of $-2.43$) on the AFMBC, but one expects posts by \texttt{Weather.com} to be related to weather forecasts, which one does not expect to see in many posts by \texttt{The Economist}. 


\section{Conclusions and Discussion}
\label{sec:dis}

Media outlets have a significant and multifaceted impact on public discourse \cite{Chaudhuri2014}. It is thus important to examine their ideological biases and heterogeneous quality levels. It can be very insightful to infer ideological positions and sentiments from the textual data of entities \cite{preoctiuc2017beyond, panda2020covid}, such as for media outlets. 
In our paper, we used a bidirectional long short-term memory (LSTM) neural network and several other machine-learning approaches to infer 
ideological biases and quality levels of media outlets based on tweets from their Twitter accounts. For both biases and qualities, we found a large correlation between the scores that we inferred from tweets and the scores in the Ad Fontes Media-Bias Chart (AFMBC). We compared a variety of ML approaches that use a bag-of-words approach with the LSTM-network approach, which incorporates word sequences, and we found that the LSTM approach outperforms the others. We thus conclude that the information from word order is a significant contributor to our natural-language-processing (NLP) task. We expect that this is also true for many other NLP tasks.

In our paper, we demonstrated that ML methods can successfully infer ideological biases and quality levels of textual data from media outlets. However, our study has several limitations. For example, we used only Twitter data and we only considered that data from a particular time window. It is essential to integrate different types of news sources, such as long articles and videos, to better infer the ideological biases and content quality of a media outlet. Additionally, we only inferred a single point in a 2D (bias, quality) space for each media outlet, but it is more realistic to represent the bias and quality of a media outlet using a probability distribution. It is also desirable to extend the analysis of ideological bias to multiple dimensions (e.g., with different dimensions for different political issues, such as social and economic issues) and to develop better approaches to evaluate inferred (bias, quality) coordinates and associated media-bias charts. Our paper provides a proof of concept for such studies, and it is important to explore these extensions.

\newpage


\appendix

\section{List of Media Outlets and the Number of Tweets of Each Outlet}
\label{tab:media outlets}

In \Cref{list_of_media}, we list the 65 media outlets that we use in our experiments. We also indicate the number of tweets from each media outlet. The tweets were collected by Littman et al.~\cite{DVN/2FIFLH_2017} between 4 August 2016 and 12 May 2020. In \Cref{sec:data}, we described our process of manually selecting a Twitter account for each media outlet. In \Cref{list_of_media}, we also give each media outlet's bias and quality scores from Ad Fontes \cite{otero2019media}.

\begin{center}
\begin{longtable}{|l|r|r|r|}

\hline
Outlet  &       Ideological Bias &    Content Quality &   Number of Tweets \\ \hline
ABC                       &  $-1.85$ &    49.87 &  39243 \\
AP                        &  $-1.06$ &    52.19 &  35132 \\
Axios                     &  $-5.74$ &    47.30 &  10647 \\
BBC                       &  $-3.03$ &    46.27 &   5685 \\
Bloomberg                 &  $-0.85$ &    47.52 &    148 \\
Breitbart                 &  18.99 &    30.64 &   9335 \\
Business Insider          &  $-0.38$ &    43.28 &  58868 \\
BuzzFeed                  &  $-7.06$ &    43.17 &  16214 \\
CBS                       &  $-1.85$ &    46.84 &  39004 \\
CNN                       &  $-8.55$ &    40.49 &  68014 \\
CNSNews                   &  25.75 &    27.75 &   5570 \\
Christian Science Monitor &  $-0.21$ &    44.27 &  24660 \\
Daily Beast               & $-12.04$ &    38.80 &  18088 \\
Daily Caller              &  20.06 &    28.80 &  25948 \\
Daily Signal              &  19.97 &    30.41 &   7812 \\
Democracy Now             & $-16.71$ &    37.54 &   8132 \\
Financial Times           &   0.62 &    47.47 &  14966 \\
Fiscal Times              &   1.52 &    44.54 &   3250 \\
Forbes                    &   0.20 &    39.84 &  19143 \\
Foreign Policy            &  $-1.65$ &    41.69 &  10496 \\
Fortune                   &   0.43 &    45.09 &  18971 \\
Forward                   &  $-5.69$ &    37.12 &   8247 \\
Fox News                  &  18.50 &    30.08 &  67254 \\
FreeSpeech TV             & $-18.74$ &    29.95 &  10970 \\
Huffington Post           & $-11.64$ &    40.17 &  18679 \\
IJR                       &   6.72 &    44.31 &   4867 \\
LA Times                  &  $-3.06$ &    49.09 &  29475 \\
MSNBC                     & $-10.88$ &    41.21 &  25312 \\
Marketwatch               &  $-0.54$ &    45.11 &  32766 \\
Mother Jones              & $-13.92$ &    40.32 &  11903 \\
National Public Radio     &  $-2.73$ &    50.22 &  16591 \\
National Review           &  16.23 &    30.95 &  11012 \\
New Republic              & $-12.83$ &    36.03 &   6097 \\
New York Post             &   5.15 &    42.42 &  30435 \\
New York Times            &  $-4.01$ &    47.54 &  43772 \\
NewsMax                   &   9.94 &    36.02 &   5016 \\
OZY                       &  $-5.43$ &    40.80 &   5991 \\
One America News Network  &  11.26 &    35.88 &   6128 \\
PBS                       &  $-2.37$ &    47.79 &   5046 \\
Politico                  &  $-5.24$ &    46.45 &  51742 \\
ProPublica                &  $-5.93$ &    48.14 &   1430 \\
Quartz                    &  $-3.89$ &    41.26 &  12646 \\
Reason                    &   4.12 &    38.28 &   7849 \\
Reuters                   &  $-0.95$ &    51.79 &  93120 \\
Slate                     & $-14.93$ &    34.20 &  37344 \\
Talking Points Memo       &  $-5.67$ &    42.24 &   6739 \\
The Atlantic              &  $-6.41$ &    40.59 &  10146 \\
The Blaze                 &  15.70 &    32.76 &   9091 \\
The Economist             &  $-2.43$ &    42.19 &  34927 \\
The Federalist            &  21.86 &    26.42 &   7767 \\
The Hill                  &   0.09 &    46.26 &  70065 \\
The Nation                & $-16.89$ &    33.54 &   2462 \\
The New Yorker            &  $-6.90$ &    41.83 &  13520 \\
The Week                  &  $-8.31$ &    33.98 &   9579 \\
Think Progress            & $-19.12$ &    35.85 &  10546 \\
Time                      &  $-4.35$ &    42.50 &  40979 \\
USA Today                 &  $-2.03$ &    46.12 &  24613 \\
Vanity Fair               & $-14.75$ &    35.22 &   6064 \\
Vox                       &  $-8.75$ &    42.33 &  15497 \\
Wall Street Journal       &   1.89 &    48.52 &  41903 \\
Washington Examiner       &  12.17 &    35.48 &  33592 \\
Washington Free Beacon    &  16.71 &    36.19 &  12062 \\
Washington Post           &  $-4.18$ &    44.57 &  45178 \\
Washington Times          &  12.97 &    37.23 &  26358 \\
Weather.com               &  $-2.43$ &    51.30 &  14345 \\\hline
\caption{The published bias and quality scores of the media outlets in the Ad Fontes Media-Bias Chart (version 5.1) and the number of tweets of each media outlet.}
\label{list_of_media}\\
\end{longtable}
\end{center}


\begin{backmatter}


\section*{Acknowledgements}

MAP acknowledges financial support from the National Science Foundation (grant number 1922952) through the Algorithms for Threat Detection (ATD) program. DN acknowledges financial support from the National Science Foundation (grant number 2011140 and 2108479) under the Division of Mathematical Sciences (DMS). 





\newcommand{\BMCxmlcomment}[1]{}

\BMCxmlcomment{

<refgrp>

<bibl id="B1">
  <title><p>Does the media matter? {A} field experiment measuring the effect of
  newspapers on voting behavior and political opinions</p></title>
  <aug>
    <au><snm>Gerber</snm><fnm>AS</fnm></au>
    <au><snm>Karlan</snm><fnm>D</fnm></au>
    <au><snm>Bergan</snm><fnm>D</fnm></au>
  </aug>
  <source>American Economic Journal: Applied Economics</source>
  <pubdate>2009</pubdate>
  <volume>1</volume>
  <issue>2</issue>
  <fpage>35</fpage>
  <lpage>-52</lpage>
</bibl>

<bibl id="B2">
  <title><p>The impact of social media on society</p></title>
  <aug>
    <au><snm>Amedie</snm><fnm>J</fnm></au>
  </aug>
  <source>Pop Culture Intersections</source>
  <pubdate>2015</pubdate>
  <volume>2</volume>
  <note>Available at \url{https://scholarcommons.scu.edu/engl_176/2}</note>
</bibl>

<bibl id="B3">
  <title><p>Information flow estimation: {A} study of news on
  {T}witter</p></title>
  <aug>
    <au><snm>South</snm><fnm>T</fnm></au>
    <au><snm>Smart</snm><fnm>B</fnm></au>
    <au><snm>Roughan</snm><fnm>M</fnm></au>
    <au><snm>Mitchell</snm><fnm>L</fnm></au>
  </aug>
  <source>Online Social Networks and Media</source>
  <pubdate>2022</pubdate>
  <volume>31</volume>
  <fpage>100231</fpage>
</bibl>

<bibl id="B4">
  <title><p>Online reactions to the 2017 {`Unite the Right'} rally in
  Charlottesville: {M}easuring polarization in {T}witter networks using media
  followership</p></title>
  <aug>
    <au><snm>Tien</snm><fnm>JH</fnm></au>
    <au><snm>Eisenberg</snm><fnm>MC</fnm></au>
    <au><snm>Cherng</snm><fnm>ST</fnm></au>
    <au><snm>Porter</snm><fnm>MA</fnm></au>
  </aug>
  <source>Applied Network Science</source>
  <pubdate>2020</pubdate>
  <volume>5</volume>
  <issue>1</issue>
  <fpage>10</fpage>
</bibl>

<bibl id="B5">
  <title><p>Pro-choice and pro-life are not enough: {An} investigation of
  abortion attitudes as a function of abortion prototypes</p></title>
  <aug>
    <au><snm>Rye</snm><fnm>B. J.</fnm></au>
    <au><snm>Underhill</snm><fnm>A</fnm></au>
  </aug>
  <source>Sexuality \& Culture</source>
  <publisher>Springer</publisher>
  <pubdate>2020</pubdate>
  <volume>24</volume>
  <fpage>1829</fpage>
  <lpage>-1851</lpage>
</bibl>

<bibl id="B6">
  <title><p>News sharing on {T}witter reveals emergent fragmentation of media
  agenda and persistent polarization</p></title>
  <aug>
    <au><snm>Cicchini</snm><fnm>T</fnm></au>
    <au><snm>Del Pozo</snm><fnm>SM</fnm></au>
    <au><snm>Tagliazucchi</snm><fnm>E</fnm></au>
    <au><snm>Balenzuela</snm><fnm>P</fnm></au>
  </aug>
  <source>European Physical Journal --- Data Science</source>
  <pubdate>2022</pubdate>
  <volume>11</volume>
  <issue>1</issue>
  <fpage>48</fpage>
</bibl>

<bibl id="B7">
  <title><p>Social media analyses for social measurement</p></title>
  <aug>
    <au><snm>Schober</snm><fnm>MF</fnm></au>
    <au><snm>Pasek</snm><fnm>J</fnm></au>
    <au><snm>Guggenheim</snm><fnm>L</fnm></au>
    <au><snm>Lampe</snm><fnm>C</fnm></au>
    <au><snm>Conrad</snm><fnm>FG</fnm></au>
  </aug>
  <source>Public Opinion Quarterly</source>
  <pubdate>2016</pubdate>
  <volume>80</volume>
  <issue>1</issue>
  <fpage>180</fpage>
  <lpage>-211</lpage>
</bibl>

<bibl id="B8">
  <title><p>Exposure to Opposing Views on Social Media can Increase Political
  Polarization</p></title>
  <aug>
    <au><snm>Bail</snm><fnm>CA</fnm></au>
    <au><snm>Argyle</snm><fnm>LP</fnm></au>
    <au><snm>Brown</snm><fnm>TW</fnm></au>
    <au><snm>Bumpus</snm><fnm>JP</fnm></au>
    <au><snm>Chen</snm><fnm>H</fnm></au>
    <au><snm>Hunzaker</snm><fnm>MF</fnm></au>
    <au><snm>Lee</snm><fnm>J</fnm></au>
    <au><snm>Mann</snm><fnm>M</fnm></au>
    <au><snm>Merhout</snm><fnm>F</fnm></au>
    <au><snm>Volfovsky</snm><fnm>A</fnm></au>
  </aug>
  <source>Proceedings of the National Academy of Sciences of the United States
  of America</source>
  <publisher>National Acad Sciences</publisher>
  <pubdate>2018</pubdate>
  <volume>115</volume>
  <issue>37</issue>
  <fpage>9216</fpage>
  <lpage>-9221</lpage>
</bibl>

<bibl id="B9">
  <title><p>Polarized democracies in comparative perspective: {T}oward a
  conceptual framework</p></title>
  <aug>
    <au><snm>McCoy</snm><fnm>J</fnm></au>
    <au><snm>Rahman</snm><fnm>T</fnm></au>
  </aug>
  <source>International Political Science Association Conference</source>
  <publisher>Poznan, Poland</publisher>
  <pubdate>2016</pubdate>
  <volume>26</volume>
  <fpage>16</fpage>
  <lpage>-42</lpage>
</bibl>

<bibl id="B10">
  <title><p>Congress: A Political-Economic History of Roll Call
  Voting</p></title>
  <aug>
    <au><snm>Poole</snm><fnm>K. T.</fnm></au>
    <au><snm>Rosenthal</snm><fnm>H.</fnm></au>
  </aug>
  <publisher>Oxford, United Kingdom: Oxford University Press</publisher>
  <pubdate>1997</pubdate>
</bibl>

<bibl id="B11">
  <title><p>The Identification of Blocs in Small Political Bodies</p></title>
  <aug>
    <au><snm>Rice</snm><fnm>S. A.</fnm></au>
  </aug>
  <source>American Political Science Review</source>
  <pubdate>1927</pubdate>
  <volume>21</volume>
  <issue>3</issue>
  <fpage>619</fpage>
  <lpage>-627</lpage>
</bibl>

<bibl id="B12">
  <title><p>A pattern analysis of the second {R}enquist {U}.{S}. {S}upreme
  {C}ourt</p></title>
  <aug>
    <au><snm>Sirovich</snm><fnm>L.</fnm></au>
  </aug>
  <source>Proceedings of the National Academy of Sciences of the United States
  of America</source>
  <pubdate>2003</pubdate>
  <volume>100</volume>
  <issue>13</issue>
  <fpage>7432</fpage>
  <lpage>-7437</lpage>
</bibl>

<bibl id="B13">
  <title><p>Social networks that matter: {T}witter under the
  microscope</p></title>
  <aug>
    <au><snm>Huberman</snm><fnm>BA</fnm></au>
    <au><snm>Romero</snm><fnm>DM</fnm></au>
    <au><snm>Wu</snm><fnm>F</fnm></au>
  </aug>
  <source>First Monday</source>
  <pubdate>2009</pubdate>
  <volume>14</volume>
  <issue>1--5</issue>
  <note>Available at \url{https://doi.org/10.5210/fm.v14i1.2317}</note>
</bibl>

<bibl id="B14">
  <title><p>Automatic detection of political opinions in tweets</p></title>
  <aug>
    <au><snm>Maynard</snm><fnm>D</fnm></au>
    <au><snm>Funk</snm><fnm>A</fnm></au>
  </aug>
  <source>Extended Semantic Web Conference</source>
  <editor>Garc{\'i}a-Castro, Ra{\'u}l and Fensel, Dieter and Antoniou,
  Grigoris</editor>
  <pubdate>2011</pubdate>
  <fpage>88</fpage>
  <lpage>-99</lpage>
</bibl>

<bibl id="B15">
  <title><p>More tweets, more votes: {S}ocial media as a quantitative indicator
  of political behavior</p></title>
  <aug>
    <au><snm>DiGrazia</snm><fnm>J</fnm></au>
    <au><snm>McKelvey</snm><fnm>K</fnm></au>
    <au><snm>Bollen</snm><fnm>J</fnm></au>
    <au><snm>Rojas</snm><fnm>F</fnm></au>
  </aug>
  <source>PloS ONE</source>
  <publisher>Public Library of Science</publisher>
  <pubdate>2013</pubdate>
  <volume>8</volume>
  <issue>11</issue>
  <fpage>e79449</fpage>
</bibl>

<bibl id="B16">
  <title><p>Beyond Binary Labels: {P}olitical Ideology Prediction of {T}witter
  Users</p></title>
  <aug>
    <au><snm>Preo{\c{t}}iuc Pietro</snm><fnm>D</fnm></au>
    <au><snm>Liu</snm><fnm>Y</fnm></au>
    <au><snm>Hopkins</snm><fnm>D</fnm></au>
    <au><snm>Ungar</snm><fnm>L</fnm></au>
  </aug>
  <source>Proceedings of the 55th Annual Meeting of the Association for
  Computational Linguistics (Volume 1: Long Papers)</source>
  <publisher>Vancouver, Canada: Association for Computational
  Linguistics</publisher>
  <pubdate>2017</pubdate>
  <fpage>729</fpage>
  <lpage>-740</lpage>
</bibl>

<bibl id="B17">
  <title><p>Stance polarity in political debates: {A} diachronic perspective of
  network homophily and conversations on {T}witter</p></title>
  <aug>
    <au><snm>Lai</snm><fnm>M</fnm></au>
    <au><snm>Tambuscio</snm><fnm>M</fnm></au>
    <au><snm>Patti</snm><fnm>V</fnm></au>
    <au><snm>Ruffo</snm><fnm>G</fnm></au>
    <au><snm>Rosso</snm><fnm>P</fnm></au>
  </aug>
  <source>Data \& Knowledge Engineering</source>
  <publisher>Elsevier</publisher>
  <pubdate>2019</pubdate>
  <volume>124</volume>
  <fpage>101738</fpage>
</bibl>

<bibl id="B18">
  <title><p>Quantifying social organization and political polarization in
  online platforms</p></title>
  <aug>
    <au><snm>Waller</snm><fnm>I</fnm></au>
    <au><snm>Anderson</snm><fnm>A</fnm></au>
  </aug>
  <source>Nature</source>
  <pubdate>2021</pubdate>
  <volume>600</volume>
  <fpage>264</fpage>
  <lpage>-268</lpage>
</bibl>

<bibl id="B19">
  <title><p>Political bias in {TV} interviews</p></title>
  <aug>
    <au><snm>Huls</snm><fnm>E</fnm></au>
    <au><snm>Varwijk</snm><fnm>J</fnm></au>
  </aug>
  <source>Discourse \& Society</source>
  <pubdate>2011</pubdate>
  <volume>22</volume>
  <issue>1</issue>
  <fpage>48</fpage>
  <lpage>-65</lpage>
</bibl>

<bibl id="B20">
  <title><p>Opinion dynamics of online social network users: {A} micro-level
  analysis</p></title>
  <aug>
    <au><snm>Kozitsin</snm><fnm>IV</fnm></au>
  </aug>
  <source>The Journal of Mathematical Sociology</source>
  <publisher>Routledge</publisher>
  <pubdate>2021</pubdate>
</bibl>

<bibl id="B21">
  <title><p>Political polarization</p></title>
  <aug>
    <au><snm>Dixit</snm><fnm>AK</fnm></au>
    <au><snm>Weibull</snm><fnm>JW</fnm></au>
  </aug>
  <source>Proceedings of the National Academy of Sciences of the United States
  of America</source>
  <pubdate>2007</pubdate>
  <volume>104</volume>
  <issue>18</issue>
  <fpage>7351</fpage>
  <lpage>-7356</lpage>
</bibl>

<bibl id="B22">
  <title><p>{COVID, BLM}, and the polarization of {US} politicians on
  {T}witter</p></title>
  <aug>
    <au><snm>Panda</snm><fnm>A</fnm></au>
    <au><snm>Siddarth</snm><fnm>D</fnm></au>
    <au><snm>Pal</snm><fnm>J</fnm></au>
  </aug>
  <source>ArXiv:2008.03263</source>
  <pubdate>2020</pubdate>
</bibl>

<bibl id="B23">
  <title><p>Media and political polarization</p></title>
  <aug>
    <au><snm>Prior</snm><fnm>M</fnm></au>
  </aug>
  <source>Annual Review of Political Science</source>
  <pubdate>2013</pubdate>
  <volume>16</volume>
  <fpage>101</fpage>
  <lpage>-127</lpage>
</bibl>

<bibl id="B24">
  <title><p>The new \url{Voteview.com}: {P}reserving and continuing {Keith
  Poole’s} infrastructure for scholars, students and observers of
  {C}ongress</p></title>
  <aug>
    <au><snm>Boche</snm><fnm>A</fnm></au>
    <au><snm>Lewis</snm><fnm>JB</fnm></au>
    <au><snm>Rudkin</snm><fnm>A</fnm></au>
    <au><snm>Sonnet</snm><fnm>L</fnm></au>
  </aug>
  <source>Public Choice</source>
  <publisher>Springer</publisher>
  <pubdate>2018</pubdate>
  <volume>176</volume>
  <issue>1--2</issue>
  <fpage>17</fpage>
  <lpage>-32</lpage>
</bibl>

<bibl id="B25">
  <title><p>Predicting political ideology from digital footprints</p></title>
  <aug>
    <au><snm>Kitchener</snm><fnm>M</fnm></au>
    <au><snm>Anantharama</snm><fnm>N</fnm></au>
    <au><snm>Angus</snm><fnm>SD</fnm></au>
    <au><snm>Raschky</snm><fnm>PA</fnm></au>
  </aug>
  <source>ArXiv:2206.00397</source>
  <pubdate>2022</pubdate>
</bibl>

<bibl id="B26">
  <title><p>Detecting Political Biases of Named Entities and Hashtags on
  {T}witter</p></title>
  <aug>
    <au><snm>Xiao</snm><fnm>Z</fnm></au>
    <au><snm>Zhu</snm><fnm>J</fnm></au>
    <au><snm>Wang</snm><fnm>Y</fnm></au>
    <au><snm>Zhou</snm><fnm>P</fnm></au>
    <au><snm>Lam</snm><fnm>{Wen Hong}</fnm></au>
    <au><snm>Porter</snm><fnm>MA</fnm></au>
    <au><snm>Sun</snm><fnm>Y</fnm></au>
  </aug>
  <source>ArXiv:2209.08110</source>
  <pubdate>2022</pubdate>
</bibl>

<bibl id="B27">
  <title><p>Studying political bias via word embeddings</p></title>
  <aug>
    <au><snm>Gordon</snm><fnm>J</fnm></au>
    <au><snm>Babaeianjelodar</snm><fnm>M</fnm></au>
    <au><snm>Matthews</snm><fnm>J</fnm></au>
  </aug>
  <source>Companion Proceedings of the Web Conference 2020</source>
  <pubdate>2020</pubdate>
  <fpage>760</fpage>
  <lpage>-764</lpage>
</bibl>

<bibl id="B28">
  <title><p>Fighting misinformation on social media using crowdsourced
  judgments of news source quality</p></title>
  <aug>
    <au><snm>Pennycook</snm><fnm>G</fnm></au>
    <au><snm>Rand</snm><fnm>DG</fnm></au>
  </aug>
  <source>Proceedings of the National Academy of Sciences of the United States
  of America</source>
  <publisher>National Acad Sciences</publisher>
  <pubdate>2019</pubdate>
  <volume>116</volume>
  <issue>7</issue>
  <fpage>2521</fpage>
  <lpage>-2526</lpage>
</bibl>

<bibl id="B29">
  <title><p>Evaluating the fake news problem at the scale of the information
  ecosystem</p></title>
  <aug>
    <au><snm>Allen</snm><fnm>J</fnm></au>
    <au><snm>Howland</snm><fnm>B</fnm></au>
    <au><snm>Mobius</snm><fnm>M</fnm></au>
    <au><snm>Rothschild</snm><fnm>D</fnm></au>
    <au><snm>Watts</snm><fnm>DJ</fnm></au>
  </aug>
  <source>Science Advances</source>
  <publisher>American Association for the Advancement of Science</publisher>
  <pubdate>2020</pubdate>
  <volume>6</volume>
  <issue>14</issue>
  <fpage>eaay3539</fpage>
</bibl>

<bibl id="B30">
  <title><p>The art of natural language processing: {C}lassical, modern and
  contemporary approaches to text document classification</p></title>
  <aug>
    <au><snm>Ferrario</snm><fnm>A</fnm></au>
    <au><snm>Naegelin</snm><fnm>M</fnm></au>
  </aug>
  <pubdate>2020</pubdate>
  <note>Available at
  \url{https://papers.ssrn.com/sol3/papers.cfm?abstract_id=3547887}</note>
</bibl>

<bibl id="B31">
  <title><p>Strategies for training large scale neural network language
  models</p></title>
  <aug>
    <au><snm>Mikolov</snm><fnm>T</fnm></au>
    <au><snm>Deoras</snm><fnm>A</fnm></au>
    <au><snm>Povey</snm><fnm>D</fnm></au>
    <au><snm>Burget</snm><fnm>L</fnm></au>
    <au><snm>{\v{C}}ernocky</snm><fnm>J</fnm></au>
  </aug>
  <source>2011 IEEE Workshop on Automatic Speech Recognition \&
  Understanding</source>
  <pubdate>2011</pubdate>
  <fpage>196</fpage>
  <lpage>-201</lpage>
</bibl>

<bibl id="B32">
  <title><p>Comparative study of {CNN} and {RNN} for natural language
  processing</p></title>
  <aug>
    <au><snm>Yin</snm><fnm>W</fnm></au>
    <au><snm>Kann</snm><fnm>K</fnm></au>
    <au><snm>Yu</snm><fnm>M</fnm></au>
    <au><snm>Sch{\"u}tze</snm><fnm>H</fnm></au>
  </aug>
  <source>ArXiv:1702.01923</source>
  <pubdate>2017</pubdate>
</bibl>

<bibl id="B33">
  <title><p>Efficient processing of deep neural networks: {A} tutorial and
  survey</p></title>
  <aug>
    <au><snm>Sze</snm><fnm>V</fnm></au>
    <au><snm>Chen</snm><fnm>YH</fnm></au>
    <au><snm>Yang</snm><fnm>TJ</fnm></au>
    <au><snm>Emer</snm><fnm>JS</fnm></au>
  </aug>
  <source>Proceedings of the IEEE</source>
  <publisher>Ieee</publisher>
  <pubdate>2017</pubdate>
  <volume>105</volume>
  <issue>12</issue>
  <fpage>2295</fpage>
  <lpage>-2329</lpage>
</bibl>

<bibl id="B34">
  <title><p>Political ideology detection using recursive neural
  networks</p></title>
  <aug>
    <au><snm>Iyyer</snm><fnm>M</fnm></au>
    <au><snm>Enns</snm><fnm>P</fnm></au>
    <au><snm>Boyd Graber</snm><fnm>J</fnm></au>
    <au><snm>Resnik</snm><fnm>P</fnm></au>
  </aug>
  <source>Proceedings of the 52nd Annual Meeting of the Association for
  Computational Linguistics (Volume 1: Long Papers)</source>
  <pubdate>2014</pubdate>
  <fpage>1113</fpage>
  <lpage>-1122</lpage>
</bibl>

<bibl id="B35">
  <title><p>Recent advances in recurrent neural networks</p></title>
  <aug>
    <au><snm>Salehinejad</snm><fnm>H</fnm></au>
    <au><snm>Sankar</snm><fnm>S</fnm></au>
    <au><snm>Barfett</snm><fnm>J</fnm></au>
    <au><snm>Colak</snm><fnm>E</fnm></au>
    <au><snm>Valaee</snm><fnm>S</fnm></au>
  </aug>
  <source>ArXiv:1801.01078</source>
  <pubdate>2017</pubdate>
</bibl>

<bibl id="B36">
  <title><p>Recursive deep models for semantic compositionality over a
  sentiment treebank</p></title>
  <aug>
    <au><snm>Socher</snm><fnm>R</fnm></au>
    <au><snm>Perelygin</snm><fnm>A</fnm></au>
    <au><snm>Wu</snm><fnm>J</fnm></au>
    <au><snm>Chuang</snm><fnm>J</fnm></au>
    <au><snm>Manning</snm><fnm>CD</fnm></au>
    <au><snm>Ng</snm><fnm>AY</fnm></au>
    <au><snm>Potts</snm><fnm>C</fnm></au>
  </aug>
  <source>Proceedings of the 2013 Conference on Empirical Methods in Natural
  Language Processing</source>
  <pubdate>2013</pubdate>
  <fpage>1631</fpage>
  <lpage>-1642</lpage>
</bibl>

<bibl id="B37">
  <title><p>Interactive Media Bias Chart</p></title>
  <aug>
    <au><snm>Otero</snm><fnm>V</fnm></au>
  </aug>
  <source>Available at
  \url{https://adfontesmedia.com/interactive-media-bias-chart/} [accessed 26
  September 2020]</source>
  <pubdate>2022</pubdate>
  <note>Accessed 26 September 2020</note>
</bibl>

<bibl id="B38">
  <title><p>Media slant is contagious</p></title>
  <aug>
    <au><snm>Widmer</snm><fnm>P</fnm></au>
    <au><snm>Galletta</snm><fnm>S</fnm></au>
    <au><snm>Ash</snm><fnm>E</fnm></au>
  </aug>
  <source>ArXiv:2202.07269</source>
  <pubdate>2022</pubdate>
</bibl>

<bibl id="B39">
  <title><p>News Outlet Tweet IDs</p></title>
  <aug>
    <au><snm>Littman</snm><fnm>J</fnm></au>
    <au><snm>Wrubel</snm><fnm>L</fnm></au>
    <au><snm>Kerchner</snm><fnm>D</fnm></au>
    <au><snm>{Bromberg Gaber}</snm><fnm>Y</fnm></au>
  </aug>
  <publisher>Harvard Dataverse</publisher>
  <pubdate>2017</pubdate>
</bibl>

<bibl id="B40">
  <title><p>A model for the influence of media on the ideology of content in
  online social networks</p></title>
  <aug>
    <au><snm>Brooks</snm><fnm>HZ</fnm></au>
    <au><snm>Porter</snm><fnm>MA</fnm></au>
  </aug>
  <source>Physical Review Research</source>
  <pubdate>2020</pubdate>
  <volume>2</volume>
  <fpage>023041</fpage>
</bibl>

<bibl id="B41">
  <title><p>Three families of automated text analysis</p></title>
  <aug>
    <au><snm>Loon</snm><fnm>A</fnm></au>
  </aug>
  <publisher>SocArXiv</publisher>
  <pubdate>2022</pubdate>
  <note>Available at \url{osf.io/preprints/socarxiv/htnej}</note>
</bibl>

<bibl id="B42">
  <title><p>Text classification algorithms: {A} survey</p></title>
  <aug>
    <au><snm>Kowsari</snm><fnm>K</fnm></au>
    <au><snm>Jafari Meimandi</snm><fnm>K</fnm></au>
    <au><snm>Heidarysafa</snm><fnm>M</fnm></au>
    <au><snm>Mendu</snm><fnm>S</fnm></au>
    <au><snm>Barnes</snm><fnm>L</fnm></au>
    <au><snm>Brown</snm><fnm>D</fnm></au>
  </aug>
  <source>Information</source>
  <publisher>Multidisciplinary Digital Publishing Institute</publisher>
  <pubdate>2019</pubdate>
  <volume>10</volume>
  <issue>4</issue>
  <fpage>150</fpage>
</bibl>

<bibl id="B43">
  <title><p>Jumping {NLP} curves: {A} review of natural language processing
  research</p></title>
  <aug>
    <au><snm>Cambria</snm><fnm>E</fnm></au>
    <au><snm>White</snm><fnm>B</fnm></au>
  </aug>
  <source>IEEE Computational Intelligence Magazine</source>
  <publisher>IEEE</publisher>
  <pubdate>2014</pubdate>
  <volume>9</volume>
  <issue>2</issue>
  <fpage>48</fpage>
  <lpage>-57</lpage>
</bibl>

<bibl id="B44">
  <title><p>Towards Enhanced Opinion Classification using {NLP}
  Techniques.</p></title>
  <aug>
    <au><snm>Bakliwal</snm><fnm>A</fnm></au>
    <au><snm>Arora</snm><fnm>P</fnm></au>
    <au><snm>Patil</snm><fnm>A</fnm></au>
    <au><snm>Varma</snm><fnm>V</fnm></au>
  </aug>
  <source>Proceedings of the Workshop on Sentiment Analysis where AI meets
  Psychology (SAAIP 2011)</source>
  <pubdate>2011</pubdate>
  <fpage>101</fpage>
  <lpage>-107</lpage>
</bibl>

<bibl id="B45">
  <title><p>Performance analysis of ensemble methods on {T}witter sentiment
  analysis using {NLP} techniques</p></title>
  <aug>
    <au><snm>Kanakaraj</snm><fnm>M</fnm></au>
    <au><snm>Guddeti</snm><fnm>RMR</fnm></au>
  </aug>
  <source>Proceedings of the 2015 IEEE 9th International Conference on Semantic
  Computing (IEEE ICSC 2015)</source>
  <pubdate>2015</pubdate>
  <fpage>169</fpage>
  <lpage>-170</lpage>
</bibl>

<bibl id="B46">
  <title><p>Sentiment analysis of {F}acebook statuses using Naive {B}ayes
  classifier for language learning</p></title>
  <aug>
    <au><snm>Troussas</snm><fnm>C</fnm></au>
    <au><snm>Virvou</snm><fnm>M</fnm></au>
    <au><snm>Espinosa</snm><fnm>KJ</fnm></au>
    <au><snm>Llaguno</snm><fnm>K</fnm></au>
    <au><snm>Caro</snm><fnm>J</fnm></au>
  </aug>
  <source>IISA 2013</source>
  <pubdate>2013</pubdate>
  <fpage>1</fpage>
  <lpage>-6</lpage>
</bibl>

<bibl id="B47">
  <title><p>Classification of tweets data based on polarity using improved
  {RBF} kernel of {SVM}</p></title>
  <aug>
    <au><snm>Gopi</snm><fnm>AP</fnm></au>
    <au><snm>Jyothi</snm><fnm>RNS</fnm></au>
    <au><snm>Narayana</snm><fnm>VL</fnm></au>
    <au><snm>Sandeep</snm><fnm>KS</fnm></au>
  </aug>
  <source>International Journal of Information Technology</source>
  <pubdate>2020</pubdate>
  <note>Available at \url{https://doi.org/10.1007/s41870-019-00409-4}</note>
</bibl>

<bibl id="B48">
  <title><p>Recurrent neural network based language model</p></title>
  <aug>
    <au><snm>Mikolov</snm><fnm>T</fnm></au>
    <au><snm>Karafi{\'a}t</snm><fnm>M</fnm></au>
    <au><snm>Burget</snm><fnm>L</fnm></au>
    <au><snm>Cernocky</snm><fnm>J</fnm></au>
    <au><snm>Khudanpur</snm><fnm>S</fnm></au>
  </aug>
  <source>Interspeech</source>
  <publisher>Makuhari, Chiba, Japan</publisher>
  <pubdate>2010</pubdate>
  <volume>2</volume>
  <fpage>1045</fpage>
  <lpage>-1048</lpage>
</bibl>

<bibl id="B49">
  <title><p>Supervised sequence labelling with recurrent neural
  networks</p></title>
  <aug>
    <au><snm>Kawakami</snm><fnm>K</fnm></au>
  </aug>
  <source>PhD thesis</source>
  <publisher>Technical University of Munich</publisher>
  <pubdate>2008</pubdate>
</bibl>

<bibl id="B50">
  <title><p>Twitter sentiment classification using distant
  supervision</p></title>
  <aug>
    <au><snm>Go</snm><fnm>A</fnm></au>
    <au><snm>Bhayani</snm><fnm>R</fnm></au>
    <au><snm>Huang</snm><fnm>L</fnm></au>
  </aug>
  <source>CS224N Project Report, Stanford University</source>
  <pubdate>2009</pubdate>
  <note>Available at
  \url{https://www-cs.stanford.edu/people/alecmgo/papers/TwitterDistantSupervision09.pdf}</note>
</bibl>

<bibl id="B51">
  <title><p>Foundations of Statistical Natural Language Processing</p></title>
  <aug>
    <au><snm>Manning</snm><fnm>C</fnm></au>
    <au><snm>Schutze</snm><fnm>H</fnm></au>
  </aug>
  <publisher>Cambridge, MA, USA: MIT Press</publisher>
  <pubdate>1999</pubdate>
</bibl>

<bibl id="B52">
  <title><p>The optimality of naive Bayes</p></title>
  <aug>
    <au><snm>Zhang</snm><fnm>H</fnm></au>
  </aug>
  <source>Proceedings of the Seventeenth International Florida Artificial
  Intelligence Research Society Conference</source>
  <pubdate>2004</pubdate>
  <fpage>562</fpage>
  <lpage>567</lpage>
</bibl>

<bibl id="B53">
  <title><p>A review on support vector machine for data
  classification</p></title>
  <aug>
    <au><snm>Bhavsar</snm><fnm>H</fnm></au>
    <au><snm>Panchal</snm><fnm>MH</fnm></au>
  </aug>
  <source>International Journal of Advanced Research in Computer Engineering \&
  Technology (IJARCET)</source>
  <pubdate>2012</pubdate>
  <volume>1</volume>
  <issue>10</issue>
  <fpage>185</fpage>
  <lpage>-189</lpage>
</bibl>

<bibl id="B54">
  <title><p>Support vector machine versus random forest for remote sensing
  image classification: {A} meta-analysis and systematic review</p></title>
  <aug>
    <au><snm>Sheykhmousa</snm><fnm>M</fnm></au>
    <au><snm>Mahdianpari</snm><fnm>M</fnm></au>
    <au><snm>Ghanbari</snm><fnm>H</fnm></au>
    <au><snm>Mohammadimanesh</snm><fnm>F</fnm></au>
    <au><snm>Ghamisi</snm><fnm>P</fnm></au>
    <au><snm>Homayouni</snm><fnm>S</fnm></au>
  </aug>
  <source>IEEE Journal of Selected Topics in Applied Earth Observations and
  Remote Sensing</source>
  <publisher>IEEE</publisher>
  <pubdate>2020</pubdate>
  <volume>13</volume>
  <fpage>6308</fpage>
  <lpage>-6325</lpage>
</bibl>

<bibl id="B55">
  <title><p>Support vector machine-based {EMG} signal classification
  techniques: {A} review</p></title>
  <aug>
    <au><snm>Toledo Perez</snm><fnm>DC</fnm></au>
    <au><snm>Rodriguez Resendiz</snm><fnm>J</fnm></au>
    <au><snm>Gomez Loenzo</snm><fnm>RA</fnm></au>
    <au><snm>Jauregui Correa</snm><fnm>J. C.</fnm></au>
  </aug>
  <source>Applied Sciences</source>
  <publisher>Multidisciplinary Digital Publishing Institute</publisher>
  <pubdate>2019</pubdate>
  <volume>9</volume>
  <issue>20</issue>
  <fpage>4402</fpage>
</bibl>

<bibl id="B56">
  <title><p>Loss functions for preference levels: {R}egression with discrete
  ordered labels</p></title>
  <aug>
    <au><snm>Rennie</snm><fnm>JDM</fnm></au>
    <au><snm>Srebro</snm><fnm>N</fnm></au>
  </aug>
  <source>IJCAI-05 Multidisciplinary Workshop on Advances in Preference
  Handling</source>
  <pubdate>2005</pubdate>
</bibl>

<bibl id="B57">
  <title><p>Classification and Regression Trees</p></title>
  <aug>
    <au><snm>Breiman</snm><fnm>L</fnm></au>
    <au><snm>Friedman</snm><fnm>JH</fnm></au>
    <au><snm>Olshen</snm><fnm>RA</fnm></au>
    <au><snm>Stone</snm><fnm>CJ</fnm></au>
  </aug>
  <publisher>New York, NY, USA: Routledge</publisher>
  <pubdate>2017</pubdate>
</bibl>

<bibl id="B58">
  <title><p>Sentiment classification of Roman-Urdu opinions using Naive
  {B}ayesian, Decision Tree and {KNN} classification techniques</p></title>
  <aug>
    <au><snm>Bilal</snm><fnm>M</fnm></au>
    <au><snm>Israr</snm><fnm>H</fnm></au>
    <au><snm>Shahid</snm><fnm>M</fnm></au>
    <au><snm>Khan</snm><fnm>A</fnm></au>
  </aug>
  <source>Journal of King Saud University-Computer and Information
  Sciences</source>
  <publisher>Elsevier</publisher>
  <pubdate>2016</pubdate>
  <volume>28</volume>
  <issue>3</issue>
  <fpage>330</fpage>
  <lpage>-344</lpage>
</bibl>

<bibl id="B59">
  <title><p>Sentiment analysis of a document using deep learning approach and
  decision trees</p></title>
  <aug>
    <au><snm>Zharmagambetov</snm><fnm>AS</fnm></au>
    <au><snm>Pak</snm><fnm>AA</fnm></au>
  </aug>
  <source>2015 Twelve International Conference on Electronics Computer and
  Computation (ICECCO)</source>
  <pubdate>2015</pubdate>
  <fpage>1</fpage>
  <lpage>-4</lpage>
</bibl>

<bibl id="B60">
  <title><p>Deep learning in neural networks: {A}n overview</p></title>
  <aug>
    <au><snm>Schmidhuber</snm><fnm>J</fnm></au>
  </aug>
  <source>Neural Networks</source>
  <publisher>Elsevier</publisher>
  <pubdate>2015</pubdate>
  <volume>61</volume>
  <fpage>85</fpage>
  <lpage>-117</lpage>
</bibl>

<bibl id="B61">
  <title><p>Twitter brand sentiment analysis: {A} hybrid system using n-gram
  analysis and dynamic artificial neural network</p></title>
  <aug>
    <au><snm>Ghiassi</snm><fnm>M</fnm></au>
    <au><snm>Skinner</snm><fnm>J</fnm></au>
    <au><snm>Zimbra</snm><fnm>D</fnm></au>
  </aug>
  <source>Expert Systems with applications</source>
  <publisher>Elsevier</publisher>
  <pubdate>2013</pubdate>
  <volume>40</volume>
  <issue>16</issue>
  <fpage>6266</fpage>
  <lpage>-6282</lpage>
</bibl>

<bibl id="B62">
  <title><p>An artificial neural network based approach for sentiment analysis
  of opinionated text</p></title>
  <aug>
    <au><snm>Sharma</snm><fnm>A</fnm></au>
    <au><snm>Dey</snm><fnm>S</fnm></au>
  </aug>
  <source>Proceedings of the 2012 ACM Research in Applied Computation
  Symposium</source>
  <pubdate>2012</pubdate>
  <fpage>37</fpage>
  <lpage>-42</lpage>
</bibl>

<bibl id="B63">
  <title><p>A document-level sentiment analysis approach using artificial
  neural network and sentiment lexicons</p></title>
  <aug>
    <au><snm>Sharma</snm><fnm>A</fnm></au>
    <au><snm>Dey</snm><fnm>S</fnm></au>
  </aug>
  <source>ACM SIGAPP Applied Computing Review</source>
  <publisher>ACM New York, NY, USA</publisher>
  <pubdate>2012</pubdate>
  <volume>12</volume>
  <issue>4</issue>
  <fpage>67</fpage>
  <lpage>-75</lpage>
</bibl>

<bibl id="B64">
  <title><p>Attention-based {LSTM} for aspect-level sentiment
  classification</p></title>
  <aug>
    <au><snm>Wang</snm><fnm>Y</fnm></au>
    <au><snm>Huang</snm><fnm>M</fnm></au>
    <au><snm>Zhu</snm><fnm>X</fnm></au>
    <au><snm>Zhao</snm><fnm>L</fnm></au>
  </aug>
  <source>Proceedings of the 2016 Conference on Empirical Methods in Natural
  Language Processing</source>
  <pubdate>2016</pubdate>
  <fpage>606</fpage>
  <lpage>-615</lpage>
</bibl>

<bibl id="B65">
  <title><p>Recent Trends in Deep Learning Based Natural Language Processing
  [Review Article]</p></title>
  <aug>
    <au><snm>Young</snm><fnm>T</fnm></au>
    <au><snm>Hazarika</snm><fnm>D</fnm></au>
    <au><snm>Poria</snm><fnm>S</fnm></au>
    <au><snm>Cambria</snm><fnm>E</fnm></au>
  </aug>
  <source>{IEEE} Computational Intelligence Magazine</source>
  <publisher>Institute of Electrical and Electronics Engineers
  ({IEEE})</publisher>
  <pubdate>2018</pubdate>
  <volume>13</volume>
  <issue>3</issue>
  <fpage>55</fpage>
  <lpage>-75</lpage>
</bibl>

<bibl id="B66">
  <title><p>Target-dependent sentiment classification with long short term
  memory</p></title>
  <aug>
    <au><snm>Tang</snm><fnm>D</fnm></au>
    <au><snm>Qin</snm><fnm>B</fnm></au>
    <au><snm>Feng</snm><fnm>X</fnm></au>
    <au><snm>Liu</snm><fnm>T</fnm></au>
  </aug>
  <source>ArXiv:1512.01100</source>
  <pubdate>2015</pubdate>
</bibl>

<bibl id="B67">
  <title><p>Hydrator</p></title>
  <aug>
    <au><cnm>DocNow</cnm></au>
  </aug>
  <source>Available at \url{https://github.com/docnow/hydrator} (accessed 26
  September 2022)</source>
  <pubdate>2020</pubdate>
</bibl>

<bibl id="B68">
  <title><p>Deep Learning with Keras</p></title>
  <aug>
    <au><snm>Gulli</snm><fnm>A.</fnm></au>
    <au><snm>Pal</snm><fnm>S.</fnm></au>
  </aug>
  <publisher>Birmingham, United Kingdom: Packt Publishing</publisher>
  <pubdate>2017</pubdate>
</bibl>

<bibl id="B69">
  <title><p>{LSTM} with sentence representations for document-level sentiment
  classification</p></title>
  <aug>
    <au><snm>Rao</snm><fnm>G</fnm></au>
    <au><snm>Huang</snm><fnm>W</fnm></au>
    <au><snm>Feng</snm><fnm>Z</fnm></au>
    <au><snm>Cong</snm><fnm>Q</fnm></au>
  </aug>
  <source>Neurocomputing</source>
  <publisher>Elsevier</publisher>
  <pubdate>2018</pubdate>
  <volume>308</volume>
  <fpage>49</fpage>
  <lpage>-57</lpage>
</bibl>

<bibl id="B70">
  <title><p>A review of recurrent neural networks: {LSTM} cells and network
  architectures</p></title>
  <aug>
    <au><snm>Yu</snm><fnm>Y</fnm></au>
    <au><snm>Si</snm><fnm>X</fnm></au>
    <au><snm>Hu</snm><fnm>C</fnm></au>
    <au><snm>Zhang</snm><fnm>J</fnm></au>
  </aug>
  <source>Neural Computation</source>
  <pubdate>2019</pubdate>
  <volume>31</volume>
  <issue>7</issue>
  <fpage>1235</fpage>
  <lpage>-1270</lpage>
</bibl>

<bibl id="B71">
  <title><p>{LSTM} network: {A} deep learning approach for short-term traffic
  forecast</p></title>
  <aug>
    <au><snm>Zhao</snm><fnm>Z</fnm></au>
    <au><snm>Chen</snm><fnm>W</fnm></au>
    <au><snm>Wu</snm><fnm>X</fnm></au>
    <au><snm>Chen</snm><fnm>PC</fnm></au>
    <au><snm>Liu</snm><fnm>J</fnm></au>
  </aug>
  <source>IET Intelligent Transport Systems</source>
  <publisher>Wiley Online Library</publisher>
  <pubdate>2017</pubdate>
  <volume>11</volume>
  <issue>2</issue>
  <fpage>68</fpage>
  <lpage>-75</lpage>
</bibl>

<bibl id="B72">
  <title><p>Densely connected convolutional networks</p></title>
  <aug>
    <au><snm>Huang</snm><fnm>G</fnm></au>
    <au><snm>Liu</snm><fnm>Z</fnm></au>
    <au><snm>Van Der Maaten</snm><fnm>L</fnm></au>
    <au><snm>Weinberger</snm><fnm>KQ</fnm></au>
  </aug>
  <source>Proceedings of the IEEE conference on computer vision and pattern
  recognition</source>
  <pubdate>2017</pubdate>
  <fpage>4700</fpage>
  <lpage>-4708</lpage>
</bibl>

<bibl id="B73">
  <title><p>Deep learning using rectified linear units ({ReLU})</p></title>
  <aug>
    <au><snm>Agarap</snm><fnm>AF</fnm></au>
  </aug>
  <source>ArXiv:1803.08375</source>
  <pubdate>2018</pubdate>
</bibl>

<bibl id="B74">
  <title><p>Adam: {A} method for stochastic optimization</p></title>
  <aug>
    <au><snm>Kingma</snm><fnm>DP</fnm></au>
    <au><snm>Ba</snm><fnm>J</fnm></au>
  </aug>
  <source>ArXiv:1412.6980</source>
  <pubdate>2014</pubdate>
</bibl>

<bibl id="B75">
  <title><p>Ethics in the New Media, Print Media and Visual Media
  Landscape</p></title>
  <aug>
    <au><snm>Chaudhuri</snm><fnm>M</fnm></au>
  </aug>
  <source>Journal of Global Communication</source>
  <publisher>Diva Enterprises Private Limited</publisher>
  <pubdate>2014</pubdate>
  <volume>7</volume>
  <issue>1</issue>
  <fpage>84</fpage>
  <lpage>-95</lpage>
</bibl>

</refgrp>
} 


\end{backmatter}
\end{document}